\documentclass[12pt]{article}
\pdfoutput=1
\usepackage{articlestyle}
\usepackage{natbib,bm}
\bibpunct[; ]{(}{)}{,}{a}{}{;}
 \usepackage{amstext}
\author{Yu Ryan Yue}
\usepackage{graphicx}
\usepackage{amsmath}
\usepackage{pdflscape}
\usepackage{subcaption} 
\usepackage{url}
\usepackage{color}
\usepackage{rotating}
\usepackage{array}
\usepackage{color}
\usepackage[percent]{overpic}
\usepackage{xcolor}
\usepackage{setspace}
\usepackage{titlesec}
\titleformat{\section}
	{\fontsize{14}{15}\bfseries\filcenter}
	{\thesection}
	{1em}
	{\MakeUppercase}
\titleformat{\subsection}
	{\fontsize{14}{15}\bfseries}
	{\thesubsection}
	{1em}
	{}

\newcommand{\nmathbf}{\bm}

\def\bfA{\nmathbf A}

\def\bfC{\nmathbf C}

\def\bfG{\nmathbf G}

\def\bfI{\nmathbf I}

\def\bfQ{\nmathbf Q}

\def\bfV{\nmathbf V}

\def\bfX{\nmathbf X}

\def\bfZ{\nmathbf Z}

\def\bfb{\nmathbf b}

\def\bff{\nmathbf f}

\def\bfu{\nmathbf u}
\def\bfv{\nmathbf v}
\def\bfw{\nmathbf w}

\def\bfy{\nmathbf y}

\def\bfbeta   {\nmathbf \beta}

\def\bftheta  {\nmathbf \theta}

\def\bfmu     {\nmathbf \mu}

\def\bfphi    {\nmathbf \phi}

\def\bfPsi     {\nmathbf \Psi}

\def\bfSigma  {\nmathbf \Sigma}

\newcommand{\bfzero}{{\nmathbf 0}}

\newcommand{\vareps}{\varepsilon}
\def\bfvareps{\nmathbf \varepsilon}

\def\boldfacefake#1{\kern-4pt
    \hbox{ \mathsurround=0pt
    \hbox to 0.4pt{$#1$\hss}\hbox to 0.4pt{$#1$\hss}\hbox {$#1$}}}

\newcommand{\be}{\begin{eqnarray}}
\newcommand{\ee}{\end{eqnarray}}
\newcommand{\ba}{\begin{eqnarray*}}
\newcommand{\ea}{\end{eqnarray*}}
\newcommand{\bc}{\begin{center}}
\newcommand{\ec}{\end{center}}
\newcommand{\btab}[1]{\begin{tabular}{#1}}
\newcommand{\etab}{\end{tabular}}

\newcommand{\fr}{\frac}

\newcommand{\nc}{\nonumber\\}

\newcommand{\reals}{\mbox{\rm I\kern-.20em R}}
\newcommand{\sreals}{\mbox{\small \rm I\kern-.20em R}}

\begin{document}

\def\spacingset#1{\renewcommand{\baselinestretch}%
{#1}\small\normalsize} \spacingset{1}

  \title{\bf A Bayesian General Linear Modeling Approach to Cortical Surface fMRI Data Analysis}

\author{Amanda F. Mejia$^{\rm a}$\thanks{A. Mejia and Y. Yue contributed equally to this work.}, Yu Ryan Yue$^{\rm b}$\footnotemark[1], 
David Bolin$^{\rm c}$, Finn Lindgren$^{\rm d}$\\ and Martin A. Lindquist$^{\rm e}$ 
\\\\
$^{\rm a}${\em{Indiana University, Bloomington, IN 47405}}\\ 
$^{\rm b}${\em{Baruch College, The City University of New York, New York, NY 10010}}\\ 
$^{\rm c}${\em{Gothenburg University, Gothenburg, Sweden}}\\ 
$^{\rm d}${\em{The University of Edinburgh, Edinburgh, UK}}\\
$^{\rm e}${\em{Johns Hopkins University, Baltimore, MD 21205}}
\date{}}


  \maketitle

\newpage
\begin{abstract}
Cortical surface fMRI (cs-fMRI) has recently grown in popularity versus traditional volumetric fMRI, as it allows for more meaningful spatial smoothing and is more compatible with the common assumptions of isotropy and stationarity in Bayesian spatial models.  However, as no Bayesian spatial model has been proposed for cs-fMRI data, most analyses continue to employ the classical, voxel-wise general linear model (GLM) \citep{worsley1995analysis}. Here, we propose a Bayesian GLM for cs-fMRI, which employs a class of sophisticated spatial processes to flexibly model latent activation fields.  We use integrated nested Laplacian approximation (INLA), a highly accurate and efficient Bayesian computation technique \citep{Rue:Mart:Chop:inla:2009}.  To identify regions of activation, we propose an excursions set method based on the joint posterior distribution of the latent fields, which eliminates the need for multiple comparisons correction.  Finally, we address a gap in the existing literature by proposing a novel Bayesian approach for multi-subject analysis.  The methods are validated and compared to the classical GLM through simulation studies and a motor task fMRI study from the Human Connectome Project.  The proposed Bayesian approach results in smoother activation estimates, more accurate false positive control, and increased power to detect truly active regions.
\end{abstract}
\noindent%
\spacingset{1.45}

\newpage
\section{INTRODUCTION}\label{sec:intro}
Functional magnetic resonance imaging (fMRI) is a popular noninvasive neuroimaging technique commonly used to localize regions of the brain activated by a task or stimulus \citep{lindquist08, poldrack2011handbook}.  Functional MRI indirectly measures neuronal activity through the BOLD (blood oxygenation level dependent) response, an indicator of hemodynamic changes that occur following neuronal activation \citep{lindquist2009modeling}. Traditional or \textit{volumetric} fMRI data consists of a time series of three-dimensional brain volumes, each composed of hundreds of thousands of equally sized volumetric elements (voxels).  While neuronal activity is known to occur in gray matter, volumetric fMRI data consists largely of a number of other tissue classes, including white matter and cerebral spinal fluid.

\begin{figure}
\begin{center}
\includegraphics[width=7in]{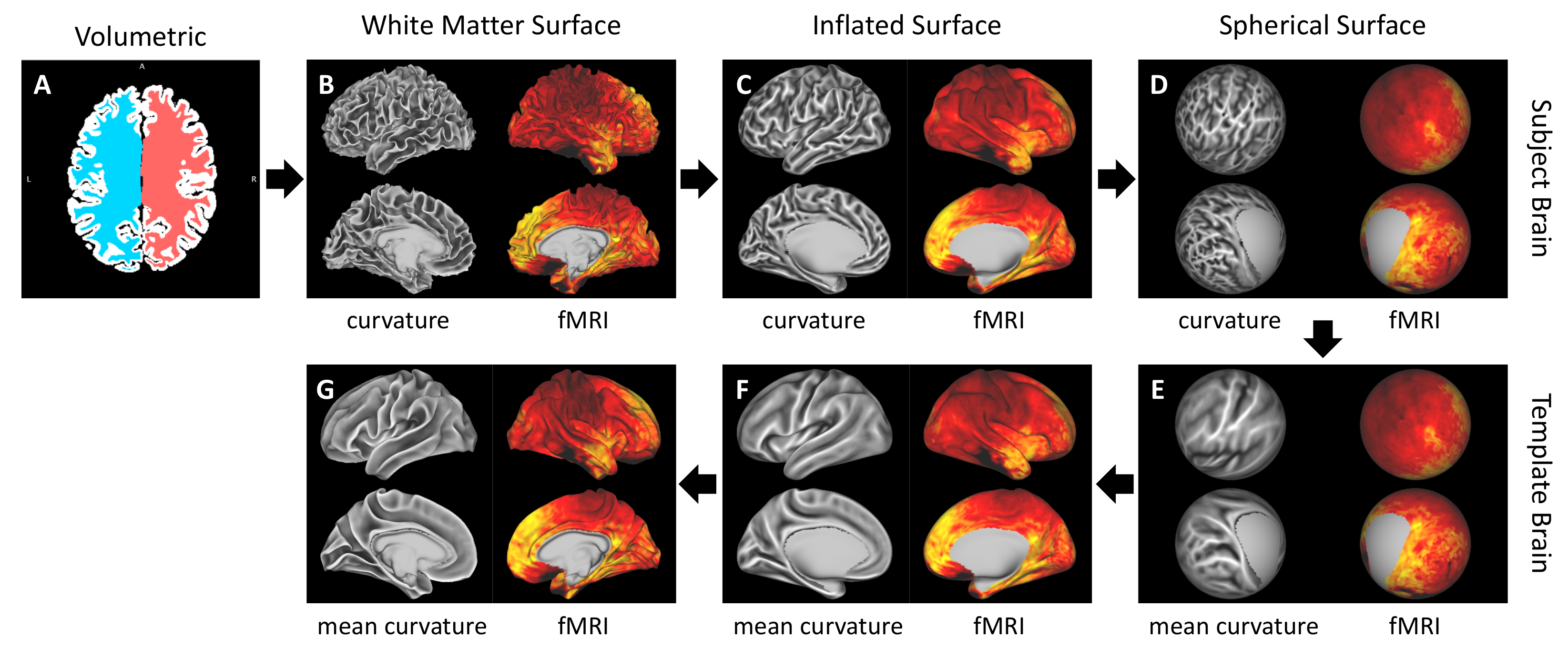}
\end{center}
\vspace{-.2in}
\singlespace \caption{\small Illustration of transformation from volumetric to surface representation.  For each four-way surface image, the top row shows the lateral (exterior) surface of each hemisphere, and the bottom rows shows the medial (interior) surface of each hemisphere. (A) shows the white matter in the left hemisphere (blue) and right hemisphere (red) and the cortical gray matter (white), as defined by the white matter and pial (not shown) boundaries.  (B) shows the white matter surface after a mesh has been applied to the white matter boundary and smoothed. (C) and (D) show two different levels of inflation of the white matter surface.  Subject brains are aligned to the template brain by aligning cortical folding patterns, indicated by curvature, on the spherical surface (D to E).  As shown in (F) and (G), the template brain can be deflated to various levels for display purposes.  In (B), (C) and (D), the curvature of the subject's brain is displayed on the left hemisphere, and the first volume of the motor task fMRI session is displayed on the right hemisphere.  In (E), (F) and (G), the average curvature of a group of subjects replaces the subject's curvature on the left hemisphere.}
\label{fig:vol_to_surf}
\end{figure}

Recently, an alternative representation of fMRI data in which the cortical gray matter is represented as a 2-dimensional manifold surface has experienced a rise in popularity \citep{fischl2012freesurfer, glasser2013minimal}.  The process of transforming volumetric to cortical surface fMRI (cs-fMRI) is illustrated in Figure \ref{fig:vol_to_surf}.  First, a high-dimension structural image is used to identify the cortical gray matter ribbon \citep{dale1999cortical}.  Second, a mesh is applied to the white matter surface, the internal boundary of the cortical gray matter, to form a 2-dimensional manifold within each hemisphere, which is then geometrically smoothed.  Third, the surface is inflated to a sphere while minimizing distance distortions, in which format subjects are registered to a standard template space by aligning anatomical folding patterns \citep{fischl1999cortical}.  Finally, the same volume-to-surface transformation is applied to each fMRI volume to obtain a cs-fMRI time series.  The resulting data is a triangular mesh consisting of approximately 30,000 vertices in each hemisphere. 


\begin{figure}
\begin{center}
\includegraphics[width=6in, trim = 0 1in 2in 1in, clip]{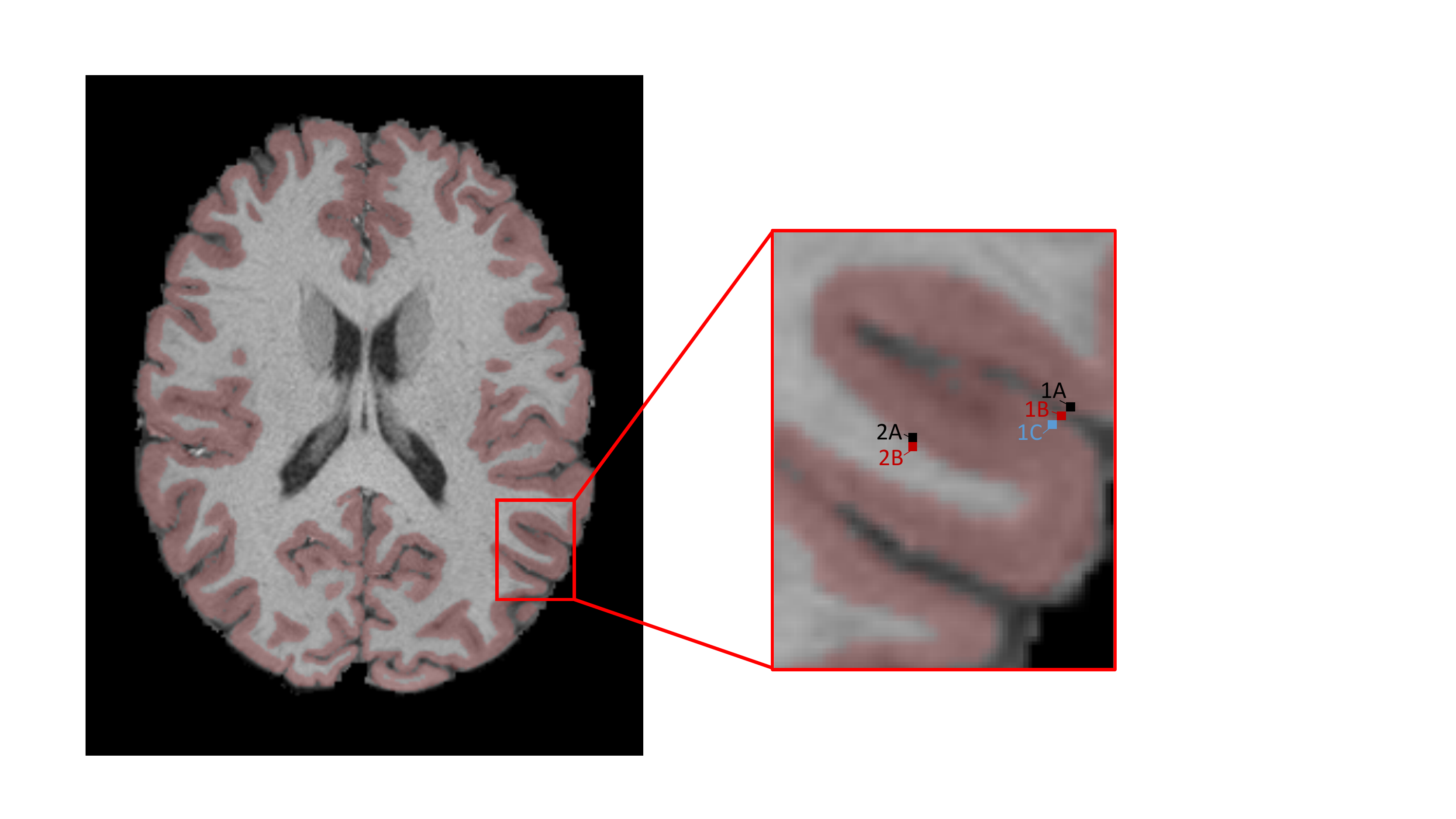}
\end{center}
\singlespace\caption{\small Distances in volumetric space.  For one subject, an axial slice of the $T_1$-weighted image is displayed, with the cortical gray matter overlaid in red.  Locations 1A, 1B and 1C are close in terms of Euclidean distance in volumetric space, but are neurologically dissimilar, as location 1A lies on one sulcal bank, location 1B lies in the cerebrospinal fluid between sulcal banks, and location 1C lies on an opposite sulcal bank.  Therefore, locations 1A and 1C may exhibit distinct task activation patterns, while location 1B would not be expected to exhibit any task-related activation.  Similarly, locations 2A and 2B are neighboring in volumetric space, but location 2A lies in the cortical gray matter while location 2B lies in the white matter and therefore would not be expected to exhibit task-related activation.  This illustrates the limitations of the volumetric representation for task fMRI analysis: the classical GLM model typically employs smoothing with a Gaussian kernel throughout the volume, which would have the result of mixing the distinct signals from locations 1A, 1B and 1C (and those from locations 2A and 2B), while a Bayesian approach assuming a stationary prior on the latent task fields would incorrectly assume the latent signal at locations 1A, 1B and 1C (and locations 2A and 2B) to be highly correlated.  By contrast, in the cortical surface representation, locations 1B and 2B would be excluded from analysis, as they do not lie within the cortical gray matter, and the latent signals at locations 1A and 1C would be assumed to have low dependence, due to the greater geodesic distance along the cortical surface between them.}
\label{fig:volumetric}
\end{figure}

Cs-fMRI offers several advantages over volumetric fMRI, including better whole-brain visualization, dimension reduction, removal of extraneous tissue types, and improved alignment of cortical areas across subjects.  However, perhaps the most important benefit of cs-fMRI is the greater neurobiological significance of distances in the cortical surface space.  Nearby locations in cs-fMRI are close in terms of distance along the cortex, and therefore tend to exhibit similar patterns of neuronal activity, while in volumetric fMRI locations that are close in terms of Euclidean distance may be neurobiologically quite dissimilar, coming from different areas of the cortex or even different tissue classes as illustrated in Figure \ref{fig:volumetric}. Thus, cs-fMRI is more appropriate for use with methods that pool information across neighbors, such as spatial smoothing or Bayesian spatial models.



The traditional task analysis method for both volumetric and cs-fMRI data is the classical {\em general linear model} (GLM), also known as the ``massive univariate'' approach, in which a linear regression model relating the observed fMRI data to the expected BOLD response to each task is fit separately at each location (e.g. voxel or vertex) in the brain \citep{friston94, hagler2006smoothing}. Prior to model fitting, the fMRI data is typically smoothed using a fixed-width Gaussian kernel in order to increase the signal-to-noise ratio (SNR) of the data. 
The coefficients of the model are estimates of the task-related activation at each location, the significance of which is tested using a $t$ or $F$ statistic.  The corresponding $p$-values are then plotted at each location to form a {\em statistical parametric map} (SPM).  To identify the areas of true activation, the null hypothesis of no activation at each location is tested by thresholding the SPM at significance level $\alpha$ \citep{worsley1995analysis}, chosen to control the family-wise error rate (FWER) or false discovery rate (FDR) at some predetermined level \citep{geno:ni:02, lindquist2015zen}. To account for correlations between tests performed at neighboring locations, popular solutions include parametric methods such as {\em random field theory} (RFT) \citep{adler:81} and nonparametric methods such as permutation tests \citep{nichols2002nonparametric}.  To avoid identifying very small regions of spurious activation, cluster-based methods first threshold the SPM at a fixed level determined by the researcher (e.g. $p=0.01$ or $p=0.001$) then use permutation tests, random field theory, or ad-hoc methods to determine significant clusters \citep{poline1993analysis}. 

While the effort to properly correct for multiple comparisons in the classical GLM is a noble and necessary one---one which is unfortunately still not universally practiced \citep{carp2012secret}---most of the traditional correction methods have been shown to suffer from various pitfalls.  In high-dimensional settings, methods that control the FWER or FDR have been shown to suffer from a lack of power to detect true effects \citep{ishwaran2003detecting, marchini04}. In the context of fMRI, parametric methods such as RFT have been found to be inaccurate due to departures of the data from the parametric assumptions \citep{nichols2003controlling, wager2009essentials, eklund2012does, eklund2016cluster}. Further, as voxels (or vertices) lack biological meaning as a unit of measure, controlling the voxel-wise (or vertex-wise) FWER is somewhat arbitrary and is sensitive to voxel size, which is gradually shrinking due to technological advances. While cluster-based methods avoid this limitation, they have been found to be sensitive to the choice of initial threshold \citep{woo2014cluster,eklund2016cluster}.  Finally, since the corrected significance threshold becomes more conservative as the number of tests increases, inference is sensitive to the size of the search volume. 

These issues are symptomatic of some of the fundamental limitations of the classical GLM, described previously by \cite{friston02} and \cite{friston03}, among others. First, while it is well-known that the activation amplitude of one voxel depends on its neighbors, the classical GLM does not account for such dependence, since the model is estimated separately at every location. 
Second, while spatial smoothing of the fMRI data prior to model fitting can increase SNR and help satisfy the assumptions of RFT, when applied to volumetric fMRI data it may also combine signal from different tissue types and across discontiguous regions of the cortex (see Figure \ref{fig:volumetric}), contaminating the signal of interest and leading to inaccurate identification of truly active regions. Smoothing of cs-fMRI data is less problematic but also tends to blur boundaries between active and non-active areas.  Furthermore, smoothed data may still fail to exhibit the Gaussian spatial autocorrelations required for RFT \citep{wald2016impacting}. Smoothing also increases dependence between tests, complicating the problem of correcting for multiple comparisons. 

To remedy these problems, several Bayesian alternatives to the classical GLM have been proposed \citep{friston03, zhang2015bayesian}. In a Bayesian GLM, specific prior distributions are assumed for the latent task activation fields and other unknown parameters in the model, and they, together with the likelihood, form a Bayesian hierarchical model. A variety of Bayesian computation techniques can be used to fit the model and obtain posterior estimates. For each location, the posterior probability that the corresponding amplitude is greater than some biologically meaningful {\em activation threshold} (often a percentage of global mean signal) is calculated, and active locations are identified by thresholding the resulting {\em posterior probability map} (PPM) at a certain level (e.g., 0.95). 
The main challenges for the successful implementation of the Bayesian GLM are selecting an appropriate prior on latent task activation fields, performing the Bayesian computation efficiently, using the \textit{joint} posterior distribution of each latent field to identify regions of activation, and performing multi-subject analysis.  We now address each of these in turn.

To account for the spatial dependence in activation levels, the model coefficients are often assumed to follow a spatial process prior, which for computational purposes should have a sparse inverse covariance structure. Several such priors have been advocated for volumetric fMRI, including a first-order Gaussian Markov random field (GMRF) prior \citep{gossl01,qui:ni:10}, Laplacian GMRF prior \citep{penny:ni:05}, sparse spatial basis function prior \citep{flandin07}, diffusion-based spatial prior \citep{harrison08}, Ising MRF prior \citep{smith07}, and nonstationary GMRF prior \citep{yue:speck:non:10, yue:loh:lind:fmri:10, yue:meta:12}.  Many priors that have been applied to volumetric fMRI are designed for data on a regular lattice and may not be applicable to cs-fMRI data, which is in the form of a triangular mesh.  In this paper, we employ a novel class of GMRF priors introduced by \cite{lind:rue:11}. The priors are obtained by solving certain stochastic partial differential equations (SPDEs) using a finite element method. Compared to the previously proposed spatial priors, the SPDE priors possess several advantages for analyzing fMRI data. First, unlike regular GMRF priors, the SPDE priors are explicit mappings of Mat\'ern Gaussian fields, which have been extensively used in statistical modeling of spatial data \citep[e.g.,][]{Guttorp:06}. Therefore, they combine the computational advantages of a GMRF with the flexibility of the Mat\'ern covariance structure in representing the dependence between locations. The Mat\'ern parametrization also provides an intuitive interpretation of the spatial properties of the random fields. Second, the SPDE priors are constructed on a flexible triangular mesh, the structure of cs-fMRI data, and can apply appropriate smoothing along boundaries.  Finally, SPDE priors are consistent under re-triangulations of the surface, which is not true in general of GMRF models that are not defined as discretisations of continuous models \citep{lind:rue:11}.  

The second challenge comes from the feasibility of Bayesian computation. Functional MRI data are often massive, consisting of 100,000-200,000 voxels, or 60,000 vertices in cs-fMRI, observed at hundreds of time points for each subject.  In our motor task fMRI study, we have $284$ time points, so that each subject's data consists of over $17$ million observations. Therefore, standard Markov chain Monte Carlo (MCMC) methods are typically too time-consuming to be practically useful for data of this size \citep[e.g.,][]{wool:ful:04}. Much attention has instead been directed towards variational Bayesian (VB) techniques \citep{woolrich2004constrained, penny:ni:05}, which achieve computational advances by assuming independence between parameters, locations, or both. However, it has been well-established that VB techniques tend to severely underestimate posterior variance in latent Gaussian models \citep{wang2005inadequacy, bishop2006pattern, Rue:Mart:Chop:inla:2009, siden2017fast}.  
We instead employ a more recently developed Bayesian inference tool based on integrated nested Laplace approximation (INLA) \citep{Rue:Mart:Chop:inla:2009}. The INLA method can directly compute accurate approximations to the posterior distributions and is able to handle large data sets by taking advantage of the sparsity of GMRFs. It is much faster than MCMC \citep{rue2016bayesian} and can be easily implemented using the \verb@R-INLA@ package (\url {http://www.r-inla.org}).

The third challenge lies in activation identification. The usual PPM approach introduced by \cite{friston03} has several limitations.  
First, as the marginal rather than the joint posterior distribution is used to determine active locations, spatial dependencies are not fully reflected in the posterior activation probabilities.  Second, using the marginal posterior distribution introduces a multiple comparisons problem, as each location is considered separately.  Finally, while each of the locations in an identified active region has marginal activation probability greater than $1-\alpha$, where $\alpha$ is a predetermined significance level, the probability that the entire region is active (i.e., all locations in the region are active) may be less than $1-\alpha$. 
Using the joint posterior distribution of each latent activation field would eliminate these issues, but doing so is typically considered computationally infeasible.  Here, we propose a computationally efficient joint PPM approach based on the excursions set method introduced by \cite{bolin2015excursion} to identify the \textit{active region}, defined as largest region such that with probability $1-\alpha$ all locations in the region are active. 

Finally, while group-level inference is often a primary goal in task fMRI studies, existing spatial Bayesian models for volumetric fMRI data are typically designed for single-subject analysis due to computational limitations \citep{siden2017fast}. To address this gap in the literature, we propose a novel approache for combining data from multiple subjects to estimate the group-level posterior distribution of each latent activation field.  Group-level regions of activation can then also be obtained using the proposed joint PPM approach. 

The remainder of this paper is organized as follows. The Bayesian GLM method is introduced in Section \ref{sec:method}, where the SPDE priors, INLA algorithm and joint PPM approach are presented. The method is then extended to multi-subject analysis in Section \ref{sec:fmri:pop}. We assess the accuracy of the proposed Bayesian methods and compare their performance to the classical GLM in a simulation study described in Section \ref{sec:simu}, followed by an application to a motor task fMRI study in Section \ref{sec:app}. We conclude with a discussion in Section \ref{sec:dis}.

\section{SINGLE SUBJECT BAYESIAN GLM}\label{sec:method}

Let $T$ be the number of time points in the fMRI timeseries and let $N$ be the number of surface vertices in each hemisphere of the brain.  For a subject and hemisphere, we have the following model:
\be\label{mod:glm:mat}
\bfy = \sum_{k=0}^K\bfX_k\bfbeta_k + \sum_{j=1}^J\bfZ_j\bfb_j + \bfvareps,\quad\bfvareps\sim N\left(\bfzero, \bfV\right).
\ee
Here $\bfy$ is an $TN\times 1$ vector containing the fMRI time series of all vertices, and the $\bfX_k$ and $\bfZ_j$ are $TN\times N$ design matrices for the activation amplitudes $\bfbeta_k$ (including baseline $\bfbeta_0$) and nuisance signals $\bfb_j$, respectively. The matrix $\bfV = \bfI_N\otimes\bfSigma(\xi,\bfphi)$, where $\bfSigma(\xi,\bfphi)$ is a $T\times T$ covariance matrix for an AR($p$) process with {\em marginal} precision $\xi$ and partial autocorrelation functions $\bfphi=(\phi_1,, \dots, \phi_p)'$ assumed for each time series, and $\otimes$ denotes the Kronecker product. For fully Bayesian inference, prior distributions are assumed on the unknown parameters in model (\ref{mod:glm:mat}). For nuisance parameters in $\bfb_j$, we take independent and diffuse Gaussian priors, that is $\bfb_j\sim N(\bfzero, \delta\bfI)$ where $\delta$ is a fixed large number. For $\xi$ and $\bfphi$, we first reparameterize them as 
\ba
\theta_1 = \log(\xi),\quad \theta_{k} = \log\left(\fr{1+\phi_{k-1}}{1-\phi_{k-1}}\right),
\ea
for $k=2,\ldots, {p+1}$. We then let $\theta_1$ follow a log gamma prior and $(\theta_2,\ldots,\theta_{p+1})$ a multivariate normal prior. The priors for $\bfbeta_k$ are described below. 


\subsection{SPDE Spatial Priors}
To account for spatial homogeneity, we need to take a spatial prior on each $\bfbeta_k$ for $k=0,\ldots, K$. A good candidate is the class of Mat\'ern Gaussian fields that has been extensively used in spatial statistics due to its flexible covariance function between locations. We say $\beta(\bfu)$ is a Mat\'ern Gaussian process if the covariance between $\bfu$ and $\bfv$ ($\bfu,\bfv\in\reals^d$) is given by
\ba
\mbox{Cov}(\bfu, \bfv) = \fr{\sigma^2}{2^{\nu-1}\Gamma(\nu)}(\kappa\|\bfu-\bfv\|)^{\nu}K_\nu(\kappa\|\bfu-\bfv\|),
\ea
where $K_\nu(\cdot)$ is the modified Bessel function of the second kind with order $\nu >0$, $\Gamma(\cdot)$ is the gamma function, $\kappa>0$ is the spatial scale, and $\sigma^2>0$ is the variance. However, a Mat\'ern spatial process is not computationally feasible for large data sets because its covariance matrix is completely dense and therefore difficult to invert. \cite{lind:rue:11} addressed this issue by deriving an explicit GMRF representation for Mat\'ern Gaussian fields through solving the following stochastic partial differential equation (SPDE) \be\label{spde:exact}
\left(\kappa^2 - \Delta\right)^{\alpha/2}\left(\tau \beta(\bfu)\right) = \mathcal{W}(\bfu),\quad\bfu\in\reals^d
\ee 
where $\Delta = \sum_{i=1}^d{\partial^2}/{\partial u_i^2}$ is the Laplacian operator, $\alpha$ is the parameter that affects the smoothness, and $\tau$ relates to the variance of $\beta$. On the right hand side of the equation, $\mathcal{W}(\bfu)$ is the Gaussian white noise process. The stationary solution $\beta$ to this SPDE is a Mat\'ern Gaussian field, and the link to the smoothness $\nu$ and variance $\sigma^2$ is $\nu=\alpha-d/2$ and $\sigma^2=\Gamma(\nu)\left[\Gamma(\alpha)(4\pi)^{d/2}\kappa^{2\nu}\tau^2\right]^{-1}$.
Spectral theory shows that an integer $\alpha$ must be chosen to obtain a Markov field. We thus let $\alpha=2$, resulting in $\nu=1$ for the smoothness of a two-dimensional field.

To obtain a Markov structure, we approximate $\beta(\bfu)$ using the following basis expansion 
\be\label{f:approx}
 \beta(\bfu) \approx\sum_{i=1}^n\psi_i(\bfu)w_i.
\ee
Here $\psi_i$ is the piecewise linear function defined on a triangular mesh, taking a value of 1 at the $i^{\text{th}}$ vertex and 0 at all other vertices; $w_i$s are the random weights that need to be estimated, and $n$ is the number of vertices in the mesh. A typical mesh is chosen to maximize the minimum interior triangle angle to ensure smooth transitions between small and large triangles. The vertices are often chosen to be the data locations, and extra vertices are added heuristically to minimize the total number of triangles needed to fulfill the size and shape constraints of the function domain. The interpretation of (\ref{f:approx}) is that the weights $\bfw = (w_1,w_2,\ldots, w_n)'$ determine the value of the field at each vertex, and the values in the interior of the triangles are determined by linear interpolation. 

The joint distribution of $\bfw$ is chosen so that the distribution of the functions $\beta(\bfu)$ approximates the distribution of solutions to the SPDE (\ref{spde:exact}). The result is that $\bfw$ is Gaussian with zero means and a sparse precision matrix given by
$$
\bfQ_{\kappa,\tau} = \tau^2\left(\kappa^4\bfC+2\kappa^2\bfG+\bfG\bfC^{-1}\bfG\right).
$$
where $\bfG$ is a sparse symmetric $n\times n$ matrix with non-zero entries in cells corresponding to neighboring locations, and $\bfC$ is a diagonal matrix \citep{bolin:13}. Consider $N$ data locations $\bfu_i ~(i=1,\ldots,N)$ and let vector $\bfbeta$ contain a realization of the random field at those locations. Then, based on (\ref{f:approx}) the SPDE prior on $\bfbeta$ is given by
\be\label{prior:f}
\bfbeta = \bfPsi\bfw,\quad\bfw\mid\kappa,\tau\sim N\left(\bfzero, \bfQ_{\kappa,\tau}^{-1}\right),
\ee
where $\bfPsi$ is the $N\times n$ sparse matrix of the basis functions. Note that $\bfPsi$ is the identity matrix if the data locations are the vertices in the mesh. Regarding the hyperparameters $\kappa$ and $\tau$, we take independent log-normal priors with zero mean and unit variance, that is $\log(\kappa)\sim N(0, 1)$ and $\log(\tau)\sim N(0, 1)$. 

As shown in \cite{lind:rue:11}, this is the best piecewise linear approximation to the continuous solution to the SPDE (\ref{spde:exact}) given a triangular mesh. Since it is a GMRF representation of Mat\'ern Gaussian fields, the SPDE prior allows us to capture both the spatial correlation and spatial smoothness that exist in a spatial process. The SPDE prior is particularly well-suited for cs-fMRI data, since the data are structured as a triangular mesh, in which connected vertices represent nearest neighbors along the cortical surface that would be expected to display similar pattens of neuronal activation.  In SPDE priors, dissimilarity between locations is related to the distance along the cortical surface, rather than Euclidean distance as in regular Mat\'ern models.  The Laplacian prior used in \cite{penny:ni:05} is a special case of the SPDE prior on a regular lattice.



\subsection{Approximate Inference by INLA}\label{sec:inla}
Based on model (\ref{mod:glm:mat}) and the priors specified in the previous sections, we may construct a Bayesian hierarchical model for our fMRI analysis as follows
\be\label{mod:bay:hier}
&&\bfy\mid\bfbeta_k,\bfb_j,\bftheta\sim N\left(\bfmu_y, \bfV\right),\quad\bfmu_y=\sum_{k=0}^K\bfX_k\bfbeta_k + \sum_{j=1}^J\bfZ_j\bfb_j,\nc
&&\bfbeta_k=\bfPsi_k\bfw_k,\quad\bfw_k\mid\bftheta\sim N\left(\bfzero, \bfQ_{\kappa_k,\tau_k}^{-1}\right),\nc
&&\bfb_j\sim N(\bfzero, \delta\bfI),\quad\bftheta\sim\pi(\bftheta),
\ee
where $\bftheta = (\xi,\phi_1,\ldots,\phi_p,\kappa_1,\ldots,\kappa_k,\tau_1,\ldots,\tau_k)$ contains all the hyperparameters, and $\pi(\bftheta)$ denotes the joint density of their hyperpriors. 

It is possible to derive the full conditional distribution of each unknown parameter, and then use MCMC-based algorithms to obtain quite a few samples from their posterior distributions and make Bayesian inferences using those samples. However, the MCMC may have mixing problems and be slow to converge, considering the size of our data and the complexity of our model. As an alternative to MCMC, \cite{Rue:Mart:Chop:inla:2009} introduced a novel Bayesian computation tool based on integrated nested Laplace approximations (INLA), which is implemented in the \verb"R-INLA" package \citep{Martins201368}. The INLA method can handle general latent Gaussian hierarchical models, including the model proposed in this paper. It accurately approximates marginal posterior densities and computes all necessary estimates faster than MCMC techniques. 

A typical latent Gaussian hierarchical model has a set of hyperparameters $\bftheta$ with prior $\pi(\bftheta)$, a set of latent Gaussian variables $\bff$ with prior $\pi(\bff|\bftheta)$ and a set of response variables $\bfy$ with likelihood $\pi(\bfy|\bff,\bftheta)$. The joint posterior distribution is then given by $\pi(\bff,\bftheta|\bfy)\propto\pi(\bfy|\bff,\bftheta)\pi(\bff|\bftheta)\pi(\bftheta)$.  The INLA method first approximates the marginal posterior of $\bftheta$ as follows:
\ba
\tilde{\pi}(\bftheta|\bfy) \propto \left.\fr{\pi(\bff,\bftheta,\bfy)}{\pi_G(\bff|\bftheta,\bfy)}\right|_{\bff=\bff^\star(\bftheta)},
\ea
where $\pi_G(\bff|\bftheta,\bfy)$ is the Gaussian approximation to the full conditional of $\bff$, and $\bff^\star(\bftheta)$ is the mode of that distribution. 
Then the approximated marginals are given by
\ba\label{eq:inla_approx}
\tilde{\pi}(\theta_j|\bfy) &=& \int\tilde{\pi}(\bftheta|\bfy)d\bftheta_{-j},\\
\tilde{\pi}(f_i | \bfy) &=& \int \tilde{\pi}(f_i|\bftheta,\bfy)\tilde{\pi}(\bftheta|\bfy)d\bftheta,
\ea
where $\bftheta_{-j}$ denotes a subvector of $\bftheta$ without element $\theta_j$. The approximated marginal of $\theta_j$ can be obtained by summing out the remaining variables $\bftheta_{-j}$ from $\tilde{\pi}(\bftheta|\bfy)$. If needed, the approximated marginal of $f_i$ is obtained by first approximating the full conditional of $f_i$ with another Laplace approximation. The parameters are then numerically integrated out  from $ \tilde{\pi}(f_i|\bftheta,\bfy)$, which gives
\be
\tilde{\pi}(f_i | \bfy) \approx \sum_\ell \lambda_\ell \tilde{\pi}(f_i|\bftheta_\ell,\bfy),
\ee
 where $\lambda_\ell$ are proportional to $\tilde{\pi}(\bftheta_\ell|\bfy)$.  The evaluation points $\bftheta_\ell$ can be chosen in different ways, depending on the relative importance of computational efficiency and accuracy in a given setting \citep{Martins201368}.

For the proposed Bayesian GLM model, given $\bff = (\bfbeta_1',\ldots,\bfbeta_K',\bfb_1,\ldots,\bfb_J')'$, $\pi(\bfy|\bff,\bftheta)$ is the Gaussian likelihood function defined in (\ref{mod:bay:hier}); $\pi(\bff|\bftheta)$ is the joint multivariate Gaussian distribution of the independent priors specified on $\bfb_j$ and $\bfbeta_k$; and $\pi(\bftheta)$ is the joint distribution of the hyperpriors. Since the likelihood is Gaussian for this model, the INLA method is greatly simplified and the only necessary approximation is the numerical integration over $\tilde\pi(\bftheta|\bfy)$.

\subsection{Joint PPM for Activation Identification}\label{sec:ppm}

After fitting model (\ref{mod:bay:hier}) with INLA, we may use the resulting estimates of activation amplitudes to identify activated brain regions. There exist a number of threshold adjustment techniques for doing this. Most of these techniques are based on first calculating the \emph{marginal} probabilities $P(f(\bfu)>\gamma)$, where $\gamma$ is an activation threshold, then defining the excedence region as $D = \{\bfu: P(f(\bfu)>\gamma)>1-\alpha\}$, where $\alpha$ is some significance level. However, the value of $\alpha$ needs to be adjusted for multiple-comparison issue, which is typically done using Type I error control, false discovery rate thresholding, or posterior probability thresholding \citep{marchini04}. In this work, we instead employ the \emph{joint} probabilities using the excursion set method introduced by \cite{bolin2015excursion}. 

We define the positive excursion set $\exset{\gamma,\alpha}{+}$ as the largest set of vertices such that with at least probability $1-\alpha$ the level $\gamma$ is exceeded at all locations in that set, which we can write as
  \begin{equation*}
\exset{\gamma,\alpha}{+}(f) = \argmax_{D}\{|D| : P(D \subseteq A_\gamma^+(f)) \geq 1-\alpha\},
\end{equation*}
where $A_\gamma^+(f) = \{ \bfu\in\Omega; f(\bfu)>\gamma \}$. Since the set is defined using the joint distribution of the random field, it should be calculated based on the posterior distribution 
\begin{equation}\label{eq:posterior}
\pi(\bff\mid\bfy) = \int \pi(\bff\mid\bfy,\mv{\theta})\pi(\mv{\theta}\mid\bfy)d\mv{\theta}.
\end{equation}
Computing  $\exset{\gamma,\alpha}{+}(f)$ based on this distribution is computationally demanding, but can be  done efficiently using the INLA technique. To do so, note that $\exset{\gamma,\alpha}{+}(f) = A_\alpha^+(F_\gamma^+)$ where $F_\gamma^+(\bfu) = \sup\{1-\alpha,\bfu\in \exset{\gamma,\alpha}{+}(f)\}$ is the so-called excursion function. Using the INLA approximation of $\pi(\bff | \bfy)$,  the excursion function is $F_\gamma^+(\bfu) =  \sum_{\ell=1}^L \lambda_\ell F_\ell(\bfu)$ where $F_\ell(\bfu)$ is the excursion function calculated for the conditional posterior $\pi(\bff | \bfy,\mv{\theta}_\ell)$ for a fixed parameter configuration $\mv{\theta}_\ell$ with corresponding weights $\lambda_\ell$, as in \eqref{eq:inla_approx}.

For the Bayesian GLM model, $\pi(\bff|\bfy,\mv{\theta}_\ell)$ is Gaussian and the computation of $F_\ell(\bfu)$, $\ell=1,\dots,L$ therefore only requires the ability to compute excursion probabilities of multivariate Gaussian distributions. This can be done efficiently using the sequential method described in \cite{bolin2015excursion}. We refer to \cite{bolin2015excursion} for further details and note that the method is implemented in the R package \verb@excursions@ \citep{bolin2016calculating}, which has an interface to \verb@R-INLA@ that greatly simplifies the usage of the method for the Bayesian GLM model.

\section{Multi-subject Bayesian GLM}\label{sec:fmri:pop}
So far, we have been concerned with the analysis of a single subject's data.  In practice, researchers often want to estimate population-level effects or pool information across a group of subjects. Multi-subject fMRI data have a hierarchical nature, with lower-level observations (e.g., individual subjects) nested within higher levels (e.g., groups of subjects). It is therefore common to use a two-level model, where in the first level there are a relatively large number of autocorrelated observations (time points), while in the second level there are a relatively small number of independent, identically distributed observations (subjects). Hierarchical models have been proposed as a way to incorporate uncertainty from subject-level models in the group-level estimates of the parameters of interest \citep{beckmann2003general, Woolrich:Neuroimage:2004, lindquist2012estimating, degras2014hierarchical}. However, as these models are typically fit separately at each voxel, they fail to account for spatial dependence in the data. On the other hand, Bayesian approaches that account for spatial dependence are usually designed for single-subject analysis. A group-level Bayesian approach that accounts for spatial dependencies for the analysis of fMRI data is therefore needed \citep{siden2017fast}.  


Theoretically, it is straightforward to combine data from all subjects to build a single model like (\ref{mod:glm:mat}) to estimate the population effects. Unfortunately, the corresponding design matrix usually has extremely large dimensions, making model fitting computationally prohibitive. Therefore, we first fit model (\ref{mod:glm:mat}) for each subject and then combine the results in a principled manner. Let us write the model for each subject $m=1,\ldots,M$ as
\ba
\bfy_m &=& \sum_{k=0}^K \bfX_{k}\bfbeta_{mk}  + \sum_{j=1}^J\bfZ_{mj}\bfb_{mj} + \bfvareps_m, \nc
&=& \bfX_m\bfbeta_m + \bfZ_m\bfb_m + \bfvareps_m, \quad \bfvareps_m\stackrel{iid}{\sim} N(\bfzero, \bfV)
\ea
where $\bfy_m$ contains all response observations for the $m^{\text{th}}$ subject, $\bfbeta_m = (\bfbeta'_{m0},\bfbeta'_{m1},\bfbeta'_{m2},\ldots,\bfbeta'_{mK})'$ and $\bfb_m = (\bfb'_{m1},\bfb'_{m2},\ldots,\bfb'_{mJ})'$ contain regression coefficients, and $\bfX = [\bfX_{0}, \bfX_{1},\bfX_{2}\ldots,\bfX_{J}]$ and $\bfZ_m = [\bfZ_{m1},\bfZ_{m2}\ldots,\bfZ_{mK}]$ are design matrices.  Since only $\bfbeta_m$ is of interest, we may regress $\bfZ_m$ out of the response $\bfy_m$ and the task design matrix $\bfX_m$ to obtain the following model
\be\label{mod:sub}
\tilde\bfy_m = \tilde\bfX_m\bfbeta_m + \bfvareps_m,\quad \bfvareps_m\sim N(\bfzero, \bfV),
\ee
where $\tilde\bfy_m$ is the fitted residual in the regression of $\bfy_m$ against $\bfZ_m$, and $\tilde\bfX_m$ is the fitted residual in the regression of $\bfX_m$ against $\bfZ_m$. 

\subsection{Two-level Modeling Approach}\label{sec:2level}

A natural approach might be to extend the familiar two-level model to a spatial Bayesian setting.  The first-level model is then the combination of the $M$ subject-level models in (\ref{mod:sub}), that is
\be\label{mod:2level:1}
\tilde\bfy = \tilde\bfX\bfbeta + \bfvareps,\quad\bfvareps\sim N(\bfzero, \bfSigma),
\ee
where $\tilde\bfy = (\tilde\bfy_1',\ldots,\tilde\bfy_M')'$, $\tilde\bfX = \mbox{diag}(\bfX_1,\ldots,\bfX_M)$, $\bfbeta = (\bfbeta_1'\ldots,\bfbeta_M')'$, $\bfvareps = (\bfvareps_1,\ldots,\bfvareps_M)'$ and $\bfSigma = \bfI_M\otimes\bfV$.  The second-level model is given by
\be\label{mod:2level:2}
\bfbeta = \bfX_G\bfbeta_G + \bfvareps_G,\quad\bfvareps_G\sim N\left(\bfzero, \sigma^2_G\bfI\right)
\ee
where $\bfX_G$ is the $NKM\times NK$ second-level design matrix, $\bfbeta_G$ is the $NK\times 1$ vector of group-level activations or contrasts, and $\sigma^2_G$ is the between-subject variance.  In the case of a single group of subjects with no covariates, we have $\bfX_G={\bf1}_M\otimes\bfI_{NK}$, but other model specifications are also possible. The two models can be collapsed as
\ba
\tilde\bfy = \tilde\bfX\bfX_G\bfbeta_G + \tilde\bfX\bfvareps_G + \bfvareps.
\ea
Unfortunately, it is computationally infeasible to fit this model using a spatial Bayesian framework. A naive solution is to first fit the $M$ subject-level models as described in Section \ref{sec:method} to obtain the posterior mean of $\bfbeta$ in (\ref{mod:2level:1}), denoted $\hat\bfbeta$. Then, use $\hat\bfbeta$ as the response variable in model (\ref{mod:2level:2}), fit this model using INLA with SPDE priors taken on the components of $\bfbeta_G$, and perform Bayesian inference on $\bfbeta_G$ as described in Section \ref{sec:method}. 

Using the posterior means, however, underestimates the uncertainty from estimating $\bfbeta$ in the subject-level models. A more accurate but computationally demanding solution is to generate posterior samples of $\bfbeta$ in model (\ref{mod:2level:1}), based on the estimated posterior distribution of $\bfbeta_m$ in each subject-level model, $m=1,\dots,M$. Let $\bfbeta^{(\ell)}$ be the $\ell^{\text{th}}$ such sample for $\ell=1,\ldots,L$.  For each sample, we use $\bfbeta^{(\ell)}$ as the response variable in model (\ref{mod:2level:2}) and fit the model using INLA. Finally, we combine the estimates from those models by averaging over the $L$ models. For example, the point estimator for $\bfbeta_G$ and the excursion function for the posterior distribution of $\bfbeta_G$ are given by
\ba
\hat\bfbeta_G = \fr{1}{L}\sum_{\ell=1}^L\hat\bfbeta_G^{(\ell)}\quad\mbox{and}\quad
F_\gamma = \fr{1}{L}\sum_{\ell=1}^LF^{(\ell)}_\gamma,
\ea 
where $\hat\bfbeta_G^{(\ell)}$ and $F^{(\ell)}_\gamma$ are the posterior mean and the excursion function based on the $\ell^{\text{th}}$ sample.  

While this sampling approach accounts for the uncertainty in the posterior distributions of the $\bfbeta_m$, it is computationally demanding due to the need to fit a separate INLA model to each of $L$ samples.  More importantly, assuming SPDE priors on both the subject-level and population-level activation fields may lead to over-smoothing in the second-level model.  Therefore, we propose the following joint modeling approach.

\subsection{Joint Modeling Approach}

For each subject $m$, $\bfbeta_m = \bfPsi_m\bfw_m$, where $\bfPsi_m = \mbox{diag}(\bfPsi_{m0},\ldots,\bfPsi_{mK})$ is block diagonal and $\bfw_m=(\bfw'_{m1},\bfw'_{m2},\ldots,\bfw'_{mK})'$.  Taking an SPDE prior on each $\bfbeta_{mk}$ implies that $\bfw_m\stackrel{i.i.d.}{\sim} N\left(\bfzero, \bfQ_\theta^{-1}\right)$ with $\bfQ_\theta=\mbox{diag}(\bfQ_{\kappa_0,\tau_0},\ldots,\bfQ_{\kappa_K,\tau_K})$ and $\bfQ_{\kappa_k,\tau_k}$ denoting the precision matrix of the SPDE prior taken on $\bfbeta_{mk}$. Then, the full conditional distribution of $\bfw_m$ is Gaussian with mean vector $\bfmu_m$ and precision matrix $\bfQ_m$, given by 
\ba
\bfmu_m = \bfQ_m^{-1}\bfPsi_m'\tilde\bfX_m'\bfV^{-1}{\tilde\bfy_m},\quad
\bfQ_m = \bfQ_{\theta} + \bfPsi_m'\tilde\bfX'_m\bfV^{-1}\tilde\bfX_m\bfPsi_m.
\ea

Since the subject-level models are independent of each other, we let $\bfw = (\bfw_1',\ldots,\bfw_M')'$ and derive the full conditional distribution of $\bfw$ to be Gaussian with mean $\bfmu = \mbox{diag}(\bfmu_1,\ldots,\bfmu_M)$ and precision matrix $\bfQ = \mbox{diag}(\bfQ_1,\ldots,\bfQ_M)$. Letting $\bfbeta = (\bfbeta_1',\ldots,\bfbeta_M')'$, we have $\bfbeta = \bfPsi\bfw$, where $\bfPsi = \mbox{diag}(\bfPsi_1,\ldots,\bfPsi_M)$. We then define the group-level $\bfbeta_G$ to be a linear combination of the subject-level $\bfbeta$, i.e., $\bfbeta_G = \bfA\bfbeta$, where $\bfA$ is a constant matrix determined by the nature of $\bfbeta_G$. For example, if we are interested in whether or not the $M$ subjects activate on average, we may define $\bfbeta_G = \sum_{m=1}^M\bfbeta_m/M$ and the corresponding $\bfA = (1/M,\ldots,1/M)\otimes\bfI_N$, an $NK\times NKM$ matrix. Since $\bfbeta_G = \bfA\bfPsi\bfw$ and the full conditional of $\bfw$ is Gaussian, it is easy to see that the full conditional of $\bfbeta_G$ is also Gaussian
\be\label{fc:beta}
\bfbeta_G\mid\bfy,\bftheta\sim N\left(\bfA\bfPsi\bfmu, \bfA\bfPsi\bfQ^{-1}\bfPsi'\bfA'\right)
\ee
where $\bftheta$ denotes the set including all the hyperparameters in $\bfmu$ and $\bfQ$.  To obtain the posterior density of $\bfbeta_G$, we need to integrate out $\bftheta$ in (\ref{fc:beta}), i.e., 
\ba
\pi(\bfbeta_G\mid\bfy)=\int\pi(\bfbeta_G\mid\bfy,\bftheta)\pi(\bftheta\mid\bfy)d\bftheta.
\ea

Unfortunately, it is hard (if not impossible) to explicitly solve the integral above. We therefore propose to numerically evaluate it using importance sampling.  The marginal posterior density is $\pi(\bftheta\mid\bfy)\propto\pi(\bftheta)\pi(\bfy\mid\bftheta)$, and since the subjects are assumed to be independent given the parameters, it is easy to show that 
\be\label{pd:theta}
\pi(\bfy\mid\bftheta)
\propto\pi(\bftheta)^{-M}\prod_{m=1}^M\pi(\bftheta\mid\bfy_m),
\ee
where $\pi(\bftheta)$ is the prior density of $\bftheta$ as specified in Section \ref{sec:method}, and $\pi(\bftheta\mid\bfy_m)$ is the posterior density of $\bftheta$ from the $m^{\text{th}}$ subject.  The proof for (\ref{pd:theta}) is given in Appendix A.  For computational ease, we approximate $\pi(\bftheta\mid\bfy_m)$ using the Gaussian approximation with mean $\tilde\bfmu_m$ and precision $\tilde\bfQ_m$ given by INLA method when fitting the subject-level models. Then, the product term in (\ref{pd:theta}) is proportional to another Gaussian density, denoted by $q(\bftheta|\bfy)$,  with precision matrix $\sum_m\tilde\bfQ_m$ and mean $(\sum_m\tilde\bfQ_m)^{-1}\sum_m\tilde\bfQ_m\tilde\bfmu_m$. As a result, the posterior distribution of $\bfbeta_G$ can be approximated by
\ba
\tilde\pi(\bfbeta_G\mid\bfy)=\int\pi(\bfbeta_G\mid\bfy,\bftheta)\pi(\bftheta)^{1-M}q(\bftheta\mid\bfy)d\bftheta.
\ea
Let $\bftheta^{(\ell)}$ be the $\ell^{\text{th}}$ sample for $\ell=1,\ldots,L$ simulated from $q(\bftheta|\bfy)$. Then, $\tilde\pi(\bfbeta_G\mid\bfy)$ can be approximated numerically as a mixture of Gaussians with weights 
$\lambda_\ell = \pi(\bftheta^{(\ell)})^{1-M}$,
and the posterior quantities of $\bfbeta_G$ can be easily computed. It's noteworthy that the highly sparse $\bfQ$ in (\ref{fc:beta}) makes the computation efficient in spite of its large dimension.  Using this approximation, we can also compute the joint PPMs as described in Section \ref{sec:ppm}. Specifically, the excursion function $F_\gamma$ for the joint posterior $\tilde\pi(\bfbeta_G\mid\bfy)$ can be written as 
\ba
F_\gamma = \fr{1}{\sum_{\ell=1}^L\lambda_\ell}\sum_{\ell=1}^L\lambda_\ell F^{(\ell)}_\gamma,
\ea 
where $F_\gamma^{(\ell)}$ is the excursion function based on the Gaussian distribution $\pi(\bfbeta_G\mid\bfy,\bftheta_\ell)$, which can be computed as described in Section \ref{sec:ppm}, and $\lambda_\ell$ are the same weights as above. Finally, the regions of activation are identified by thresholding $F_\gamma$ at a certain level, e.g., 95\%.


Since the proposed joint modeling approach is based on the full joint distribution of the population-level activation fields, it is more appropriate than the naive, two-level modeling approach.  It also makes few additional assumptions beyond the Gaussian approximation of the posterior densities $\pi(\bftheta\mid\bfy_m)$, which is likely to introduce minimal loss of accuracy in our model.  It is therefore expected to result in the most accurate estimates and regions of activation, and fitting the model is quite computationally feasible due to the sparsity of the full joint precision matrix. We compare the joint and two-level modeling approaches using experimental fMRI data in Section 5.

\section{Simulation Study}\label{sec:simu}

\begin{figure}
\begin{center}
\begin{tabular}[t]{rccc}
& Activation Profile\hspace{-0.2in} & Activation Amplitude\hspace{0.2in} & Active Regions \\
\raisebox{-\height}{\begin{picture}(0,100)
  \put(-10,45){\rotatebox[origin=c]{90}{Activation 1}}
\end{picture}} &
\raisebox{-\height}{\includegraphics[width=2.75in, trim=1cm 0 0.5cm 2cm, clip]{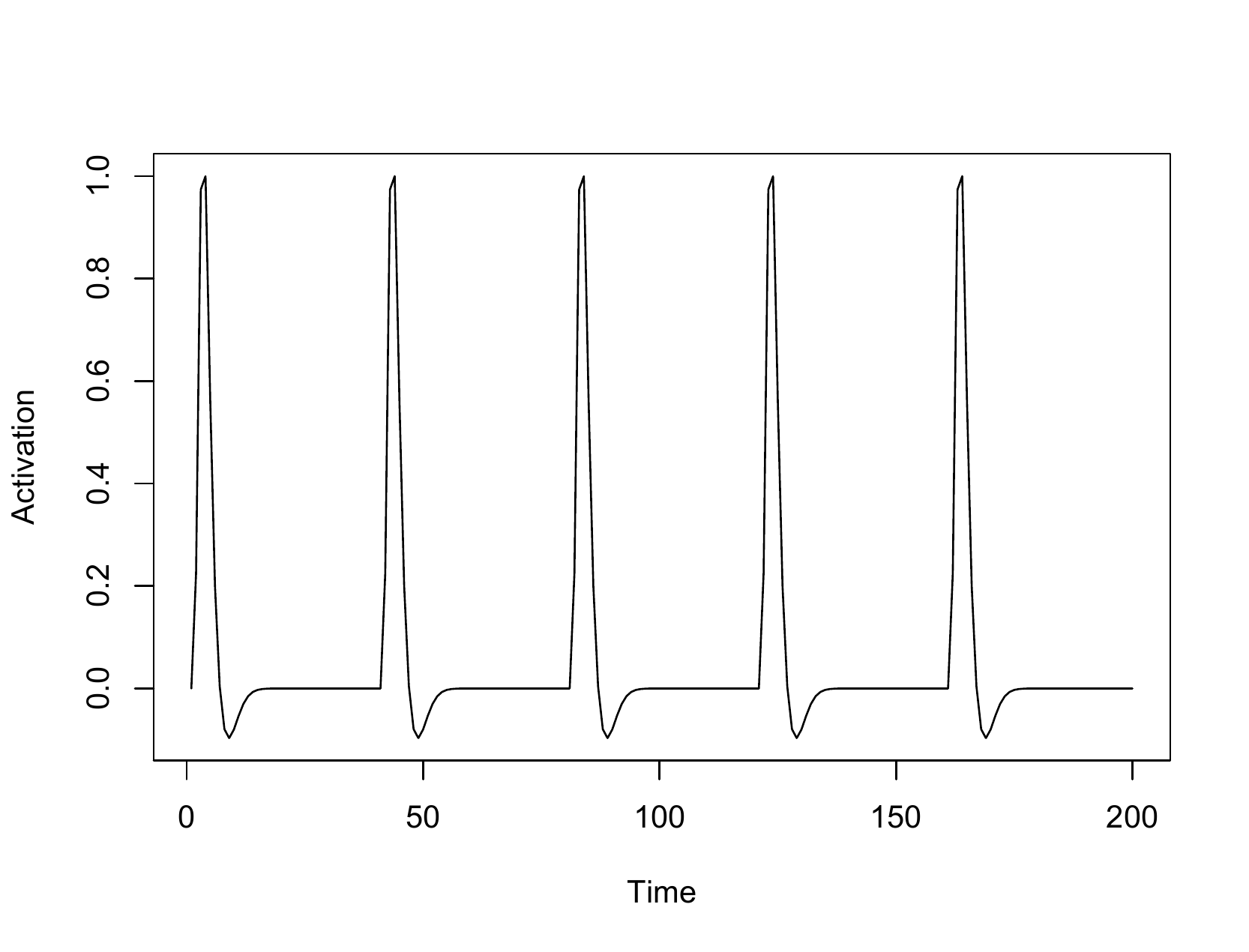}} &
\raisebox{-\height}{\includegraphics[height=1.5in, trim=2.06cm 2.58cm 0.9cm 2cm, clip]{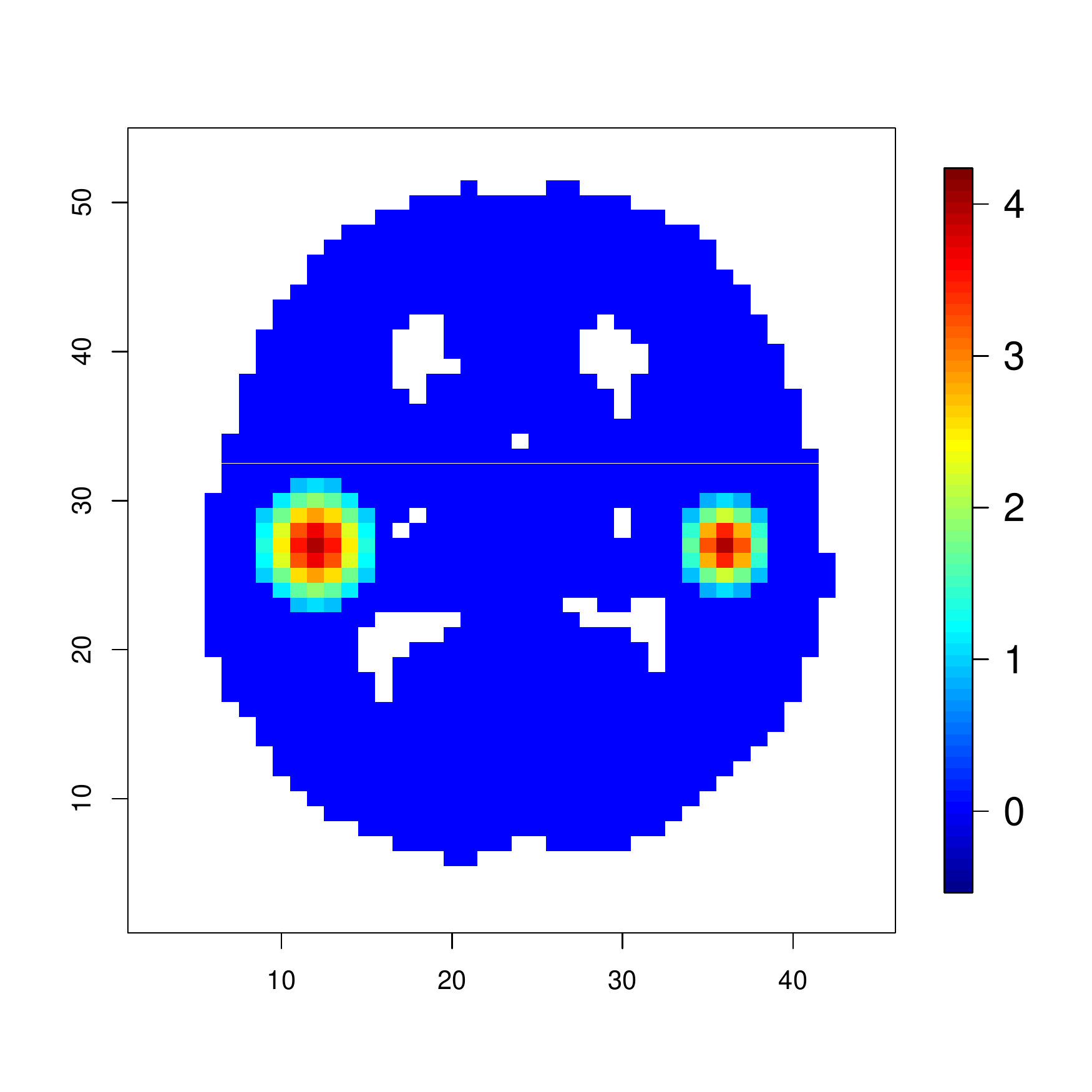}} &
\raisebox{-\height}{\includegraphics[height=1.5in, trim=2.06cm 2.58cm 3cm 2cm, clip]{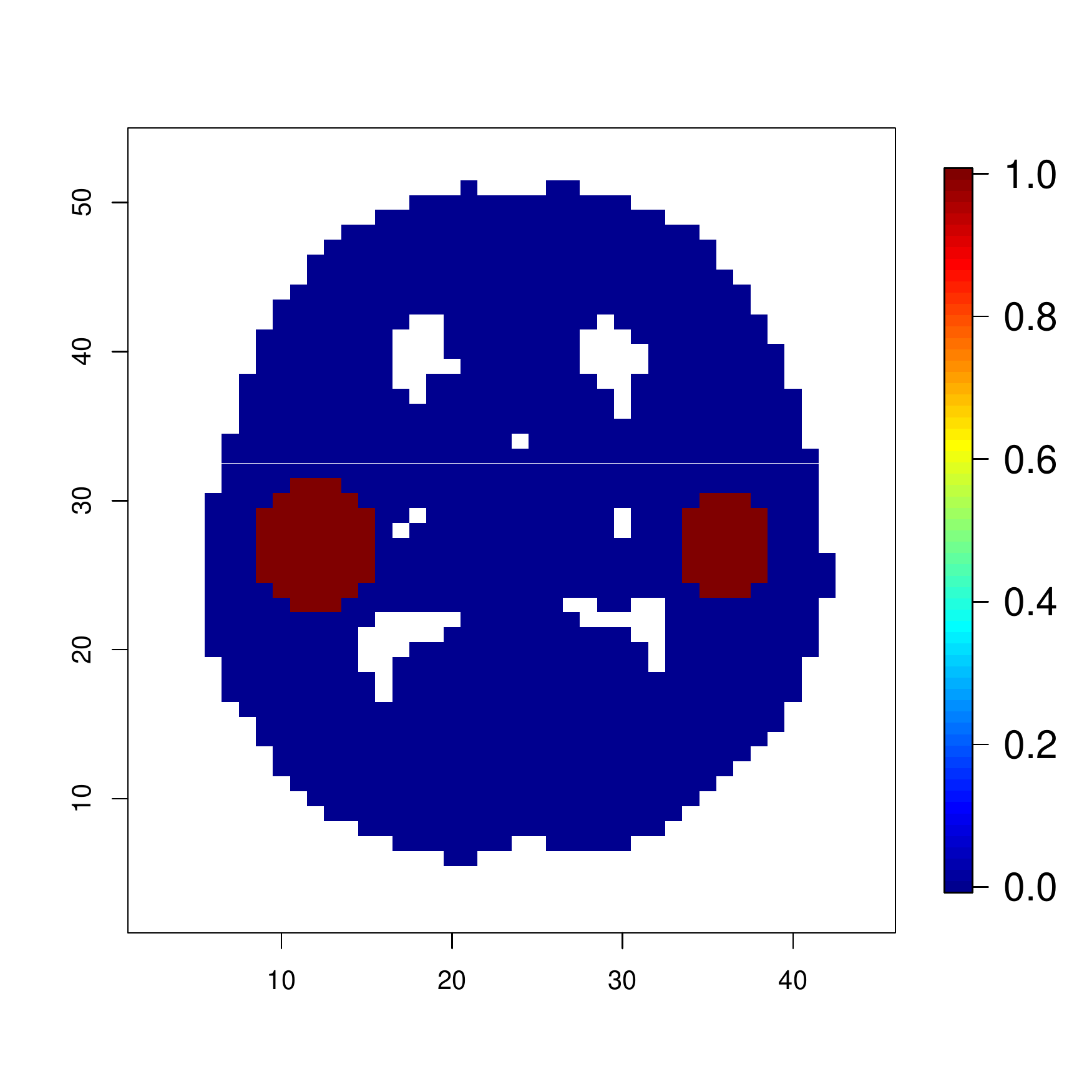}} \\[-8pt]
\raisebox{-\height}{\begin{picture}(0,100)
  \put(-10,45){\rotatebox[origin=c]{90}{Activation 2}}
\end{picture}} &
\raisebox{-\height}{\includegraphics[width=2.75in, trim=1cm 0 0.5cm 2cm, clip]{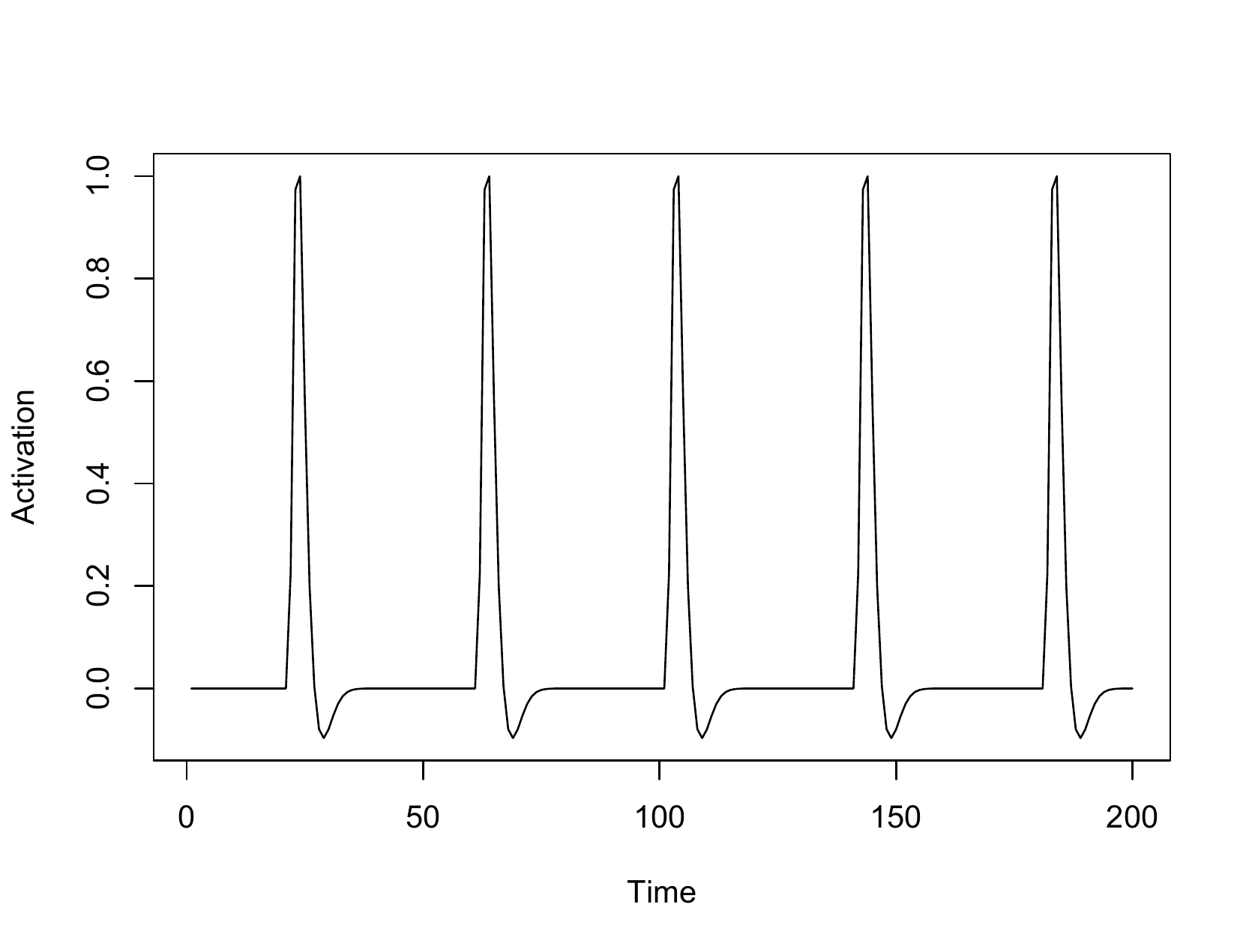}} &
\raisebox{-\height}{\includegraphics[height=1.5in, trim=2.06cm 2.58cm 0.9cm 2cm, clip]{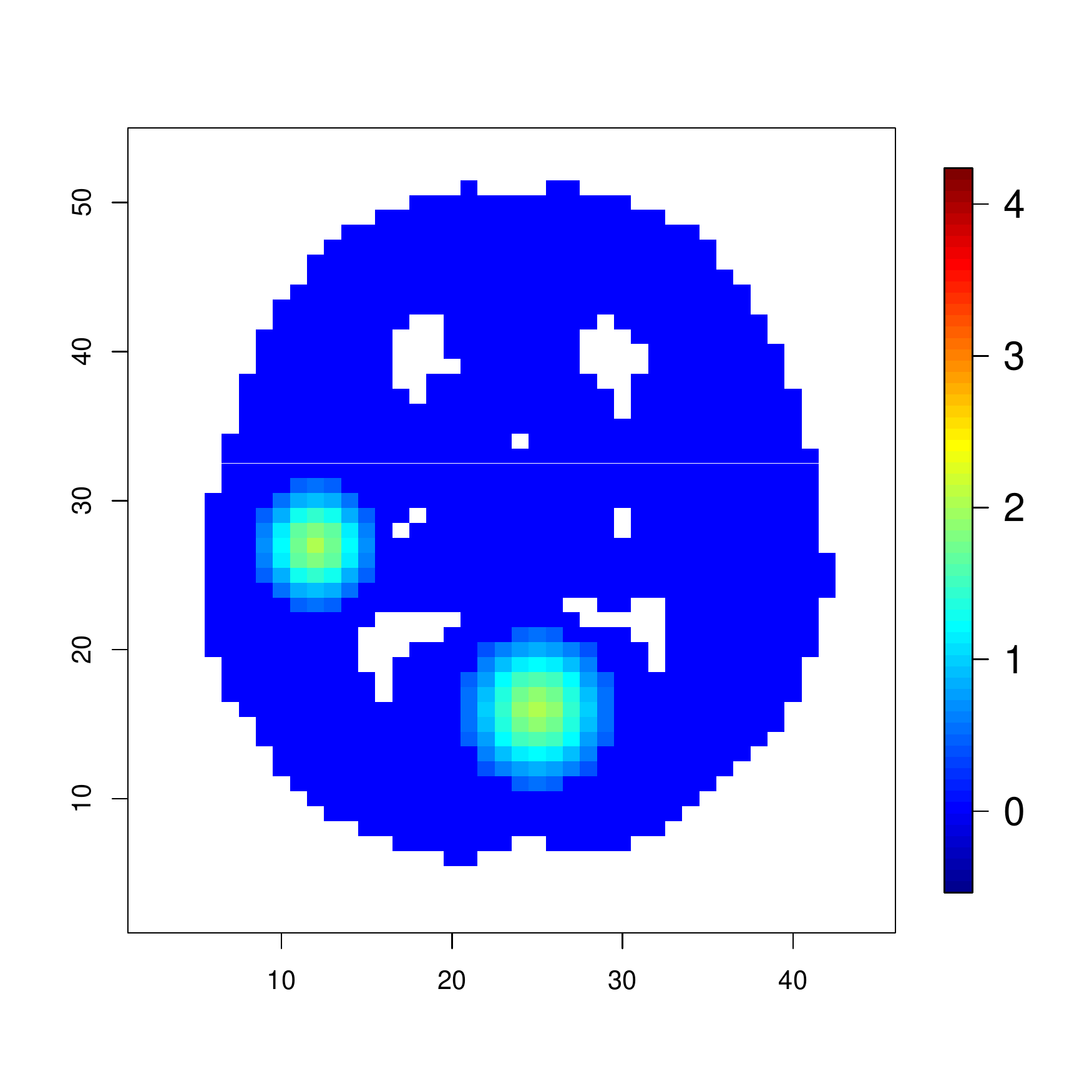}} &
\raisebox{-\height}{\includegraphics[height=1.5in, trim=2.06cm 2.58cm 3cm 2cm, clip]{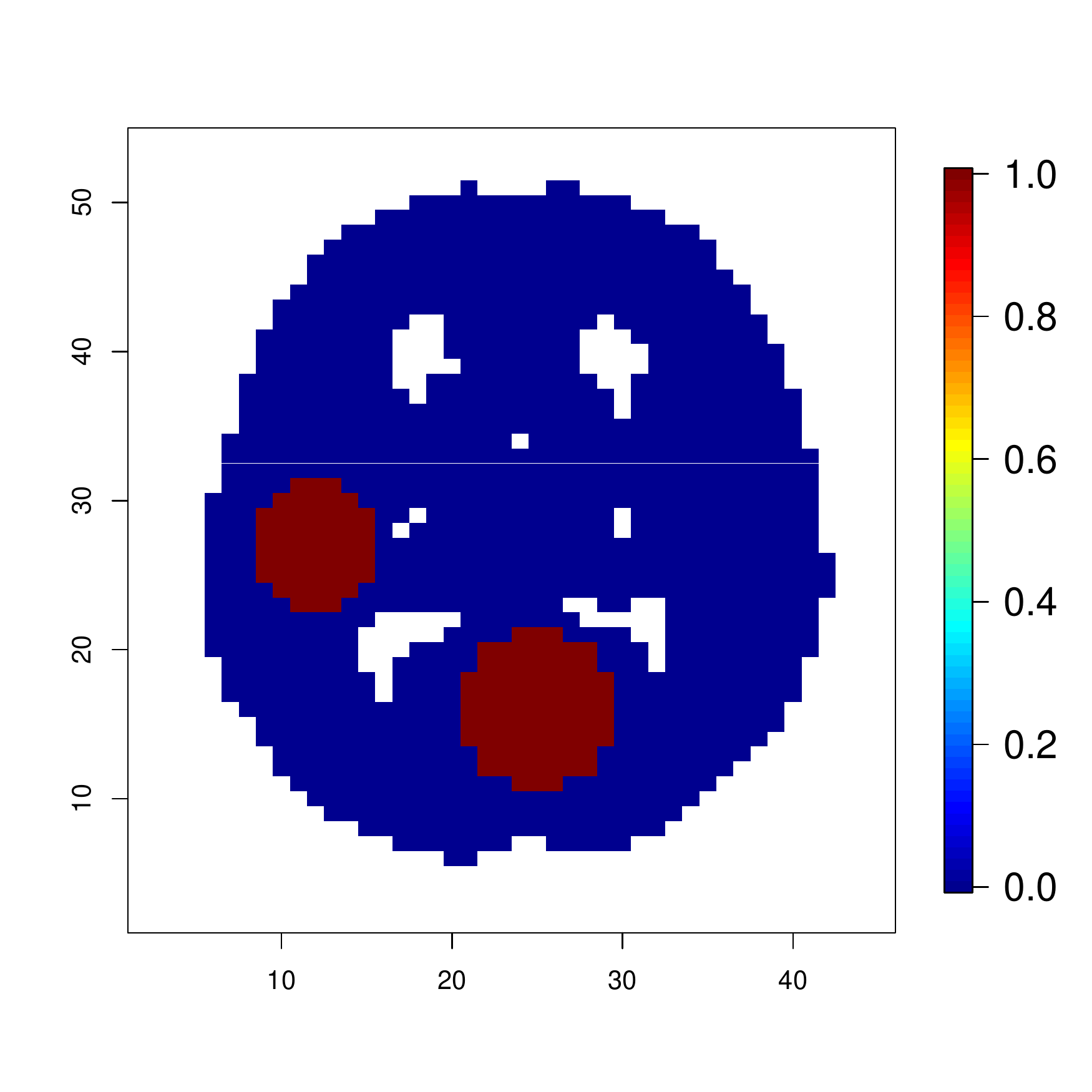}} \\[-20pt]
\end{tabular}
\end{center}
\singlespace\caption{\small Activation profiles (left), activation amplitude maps (middle), and active regions (right) for both tasks in the simulation study.}
\label{fig:act}
\end{figure}
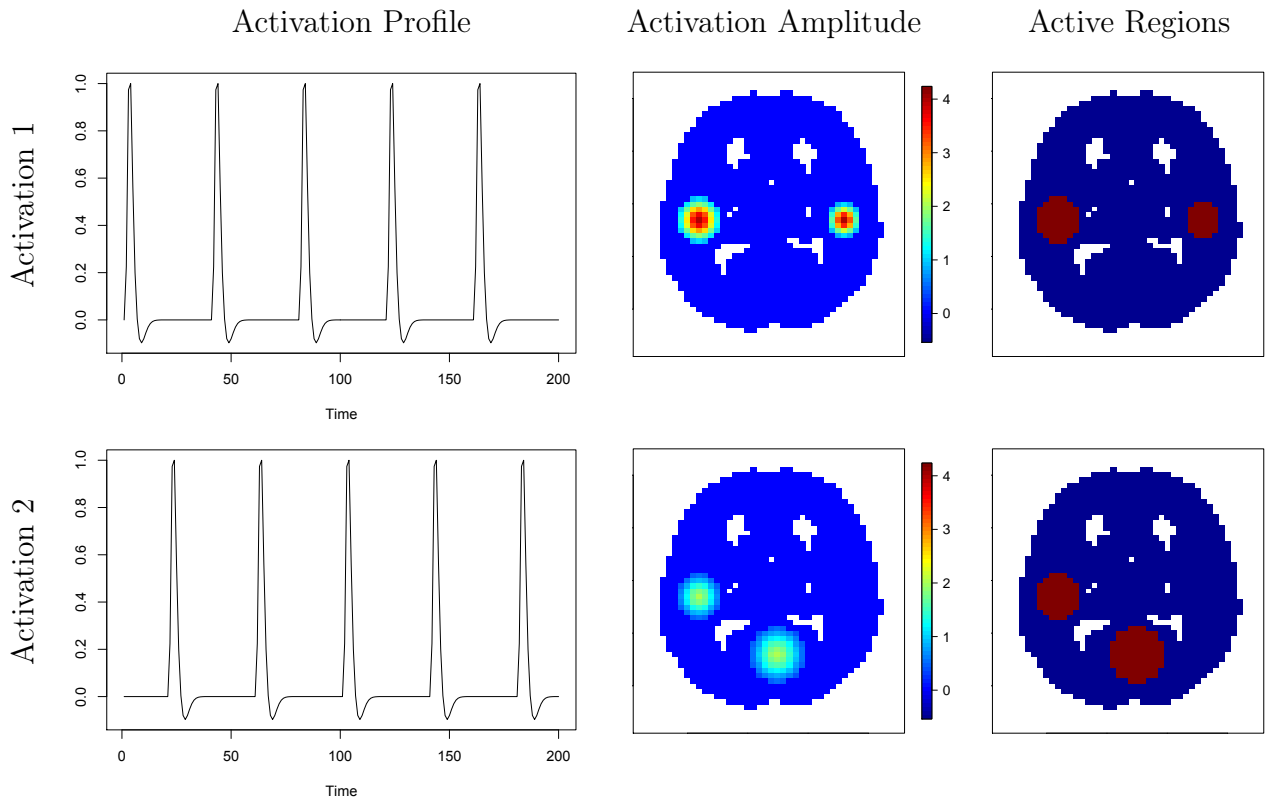

We constructed a $46 \times 55$ phantom image using a binary gray matter mask provided by the SPM 8 software package (Welcome Trust, UCL).  The selected slice contains $1256$ voxels within the brain.  The field-of-view of the image was assumed to be $192$ mm.  A dynamic image time series of length $T =200$ was simulated as follows.  Two task activation profiles $x_1$ and $x_2$, depicted in the first column of Figure \ref{fig:act}, were created by convolving a canonical hemodynamic response function \citep{friston1998event} with two task-specific binary stimulus functions.  


Activation amplitude maps $\bfbeta_1$ and $\bfbeta_2$ were formed by placing Dirac functions at three separate locations on the phantom image and smoothing each with Gaussian kernels with full width at half maximums (FWHMs) of $10$, $15$ and $20$ mm, respectively.  Smoothing was only performed here to generate spatially smooth activation fields; no smoothing was performed on the actual simulated data for the Bayesian GLM.  The signal at each voxel $v$ was generated by weighting $(x_1,x_2)$ by $(a_{1v}, a_{2v}) = (1, 0)$ for voxels within the first region, $(1, 0.5)$ within the second region, and $(0,0.5)$ within the third region.  The resulting activation amplitudes and activated regions, consisting of locations with activation amplitude greater than zero, are shown in Figure \ref{fig:act}.  The baseline field was based on the gray matter prior map, and AR(1) errors with unit variance and autocorrelation coefficient of 0.3 were independently generated at each voxel.  In summary, the data at locations $v=1,\ldots,1256$ and time points $t=1,\ldots,200$ was simulated as
\be
\label{mod:simu}
y_v(t)=\beta_{0v}+a_{1v}z_1(t)\beta_{1v}+a_{2v}z_2(t)\beta_{2v}+\vareps_v(t),
\ee
where $\beta_{0v}$ is the baseline signal, $\beta_{1v}$ and $\beta_{2v}$ are the activation amplitudes, and $\vareps_v(t)$ is random autoregressive error.  Example simulated images at two different time points are shown in Figure S1 in Appendix B.

\begin{figure}
\begin{center}
\includegraphics[width=4in, trim=1cm 3cm 0 2cm, clip]{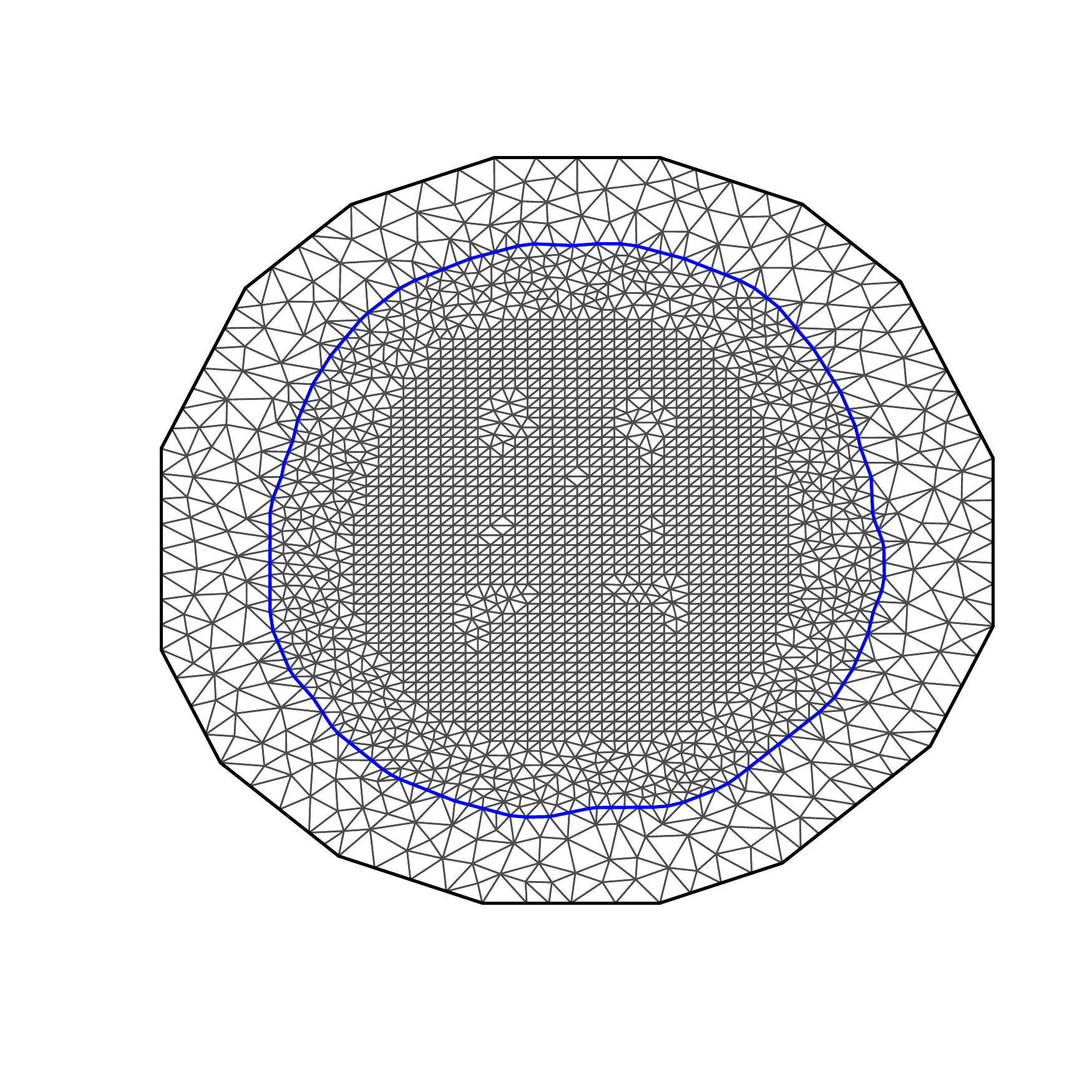}
\singlespace\caption{\small The triangular mesh used in the simulation study.}
\label{fig:mesh} 
\end{center}
\end{figure}

We fit model (\ref{mod:simu}) using the proposed Bayesian GLM method. Figure \ref{fig:mesh} displays the triangular mesh, where we see regular triangulation inside the brain and two boundary layers of larger triangles surrounding the brain and within the ventricles.  Regions of activation were identified using the joint and marginal PPM approaches with excursion level $\gamma=0$ and significance levels $\alpha=0.05$ and $0.01$.  For comparison purposes, we also applied the classical GLM method to the simulated data.  We first spatially smoothed the data using a Gaussian kernel with $6$ mm FWHM, then fit a linear regression model with AR(1) errors to each voxel separately.  The AR coefficient was estimated at each voxel by solving the Yule-Walker equations of the residuals of a linear model with uncorrelated errors, then averaging across all voxels.  Regions of activation were identified by performing a $t$-test at every voxel, accounting for multiple comparisons with FDR correction or FWER correction.  To control the FDR at $q=0.05$ and $q=0.01$, we used the Benjamini-Yekutieli procedure for dependent observations \citep{benjamini2001control}; to control the FWER at $\alpha=0.05$ and $\alpha=0.01$ we performed a permutation test by randomly reordering the prewhitened time series of all voxels $1000$ times, re-fitting the model after each reordering, to estimate the null distribution of the maximum $t$-statistic.


Model fitting was done by implementing the \verb@R-INLA@ package on a server with Intel Xeon 32-core CPU at 2.70GHz and 256GB of memory.  
Computation time to fit the model in (\ref{mod:simu}) was less than 20 minutes and would be reduced to 5 minutes if independent Gaussian errors were assumed instead of AR(1) errors.  Therefore, if the temporal pattern is not of interest, it may be removed from the data in the pre-processing stage to improve computational efficiency. This approach is adopted for our experimental data analysis, described in Section \ref{sec:app}.  Using the \verb@excursions@ package, the computation time to estimate the excursion function based on the joint PPM was approximately 30 seconds.


\begin{figure}
\begin{center}
\begin{tabular}{cccc}
& True Amplitude & Bayesian GLM & Classical GLM\hspace{0.28in} \\[8pt]
\begin{picture}(0,140)
  \put(-5,70){\rotatebox[origin=c]{90}{Activation 1}}
\end{picture} &
\includegraphics[height=2in, trim=2.06cm 2.58cm 3.1cm 2cm, clip]{beta1_true_new} &
\includegraphics[height=2in, trim=2.06cm 2.58cm 3.1cm 2cm, clip]{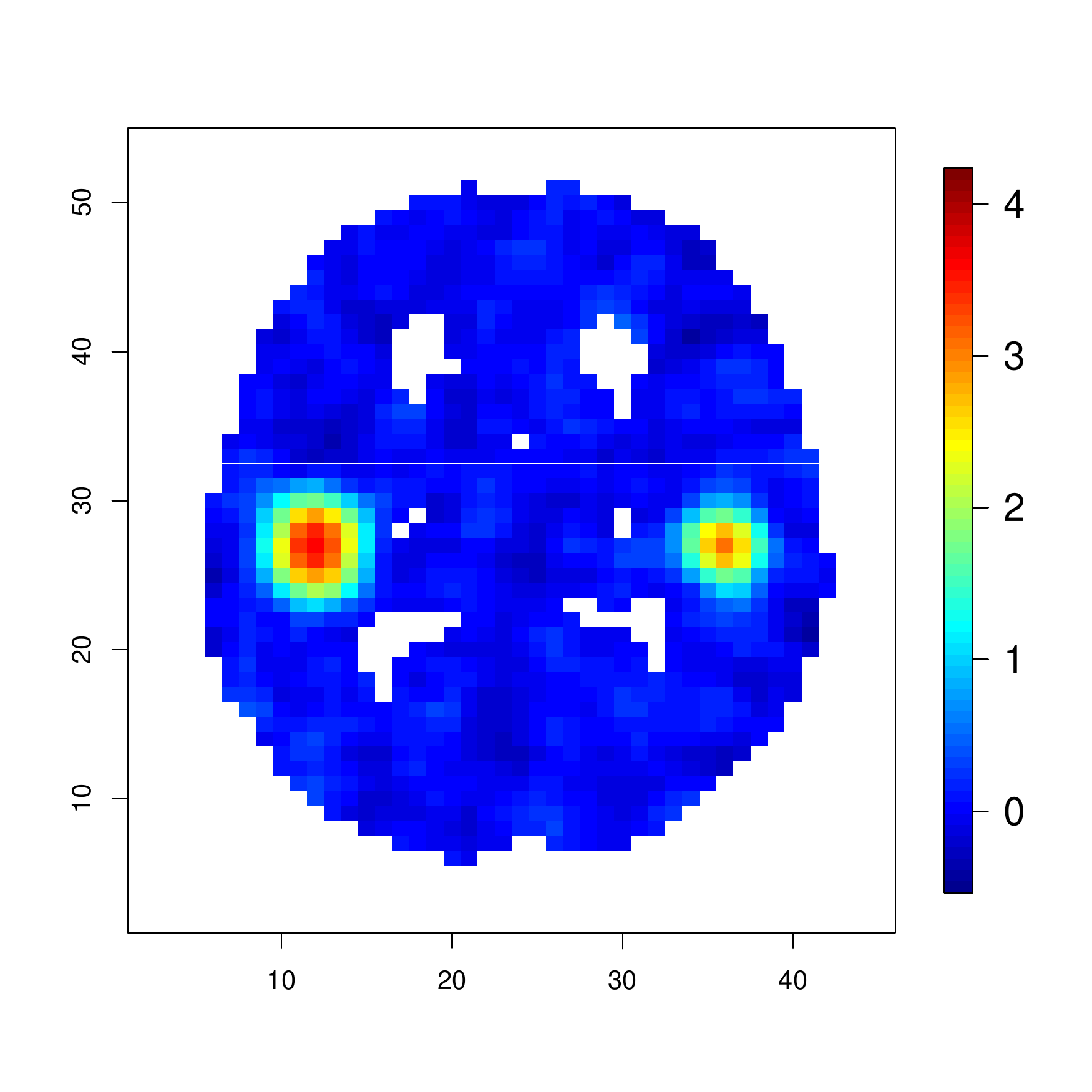} & 
\includegraphics[height=2in, trim=2.06cm 2.58cm 0.9cm 2cm, clip]{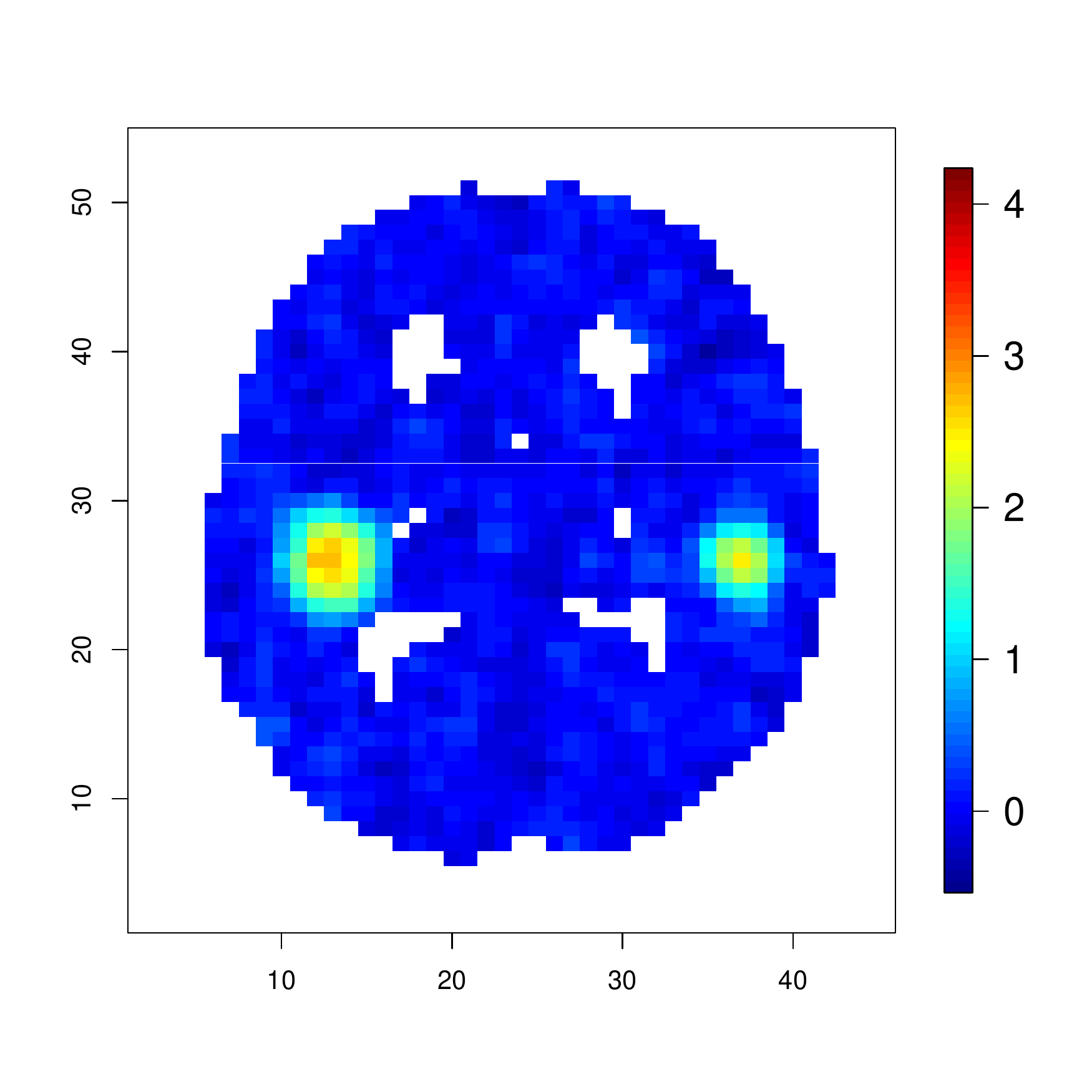} \\[8pt]
\begin{picture}(0,140)
  \put(-5,70){\rotatebox[origin=c]{90}{Activation 2}}
\end{picture} &
\includegraphics[height=2in, trim=2.06cm 2.58cm 3.1cm 2cm, clip]{beta2_true_new} &
\includegraphics[height=2in, trim=2.06cm 2.58cm 3.1cm 2cm, clip]{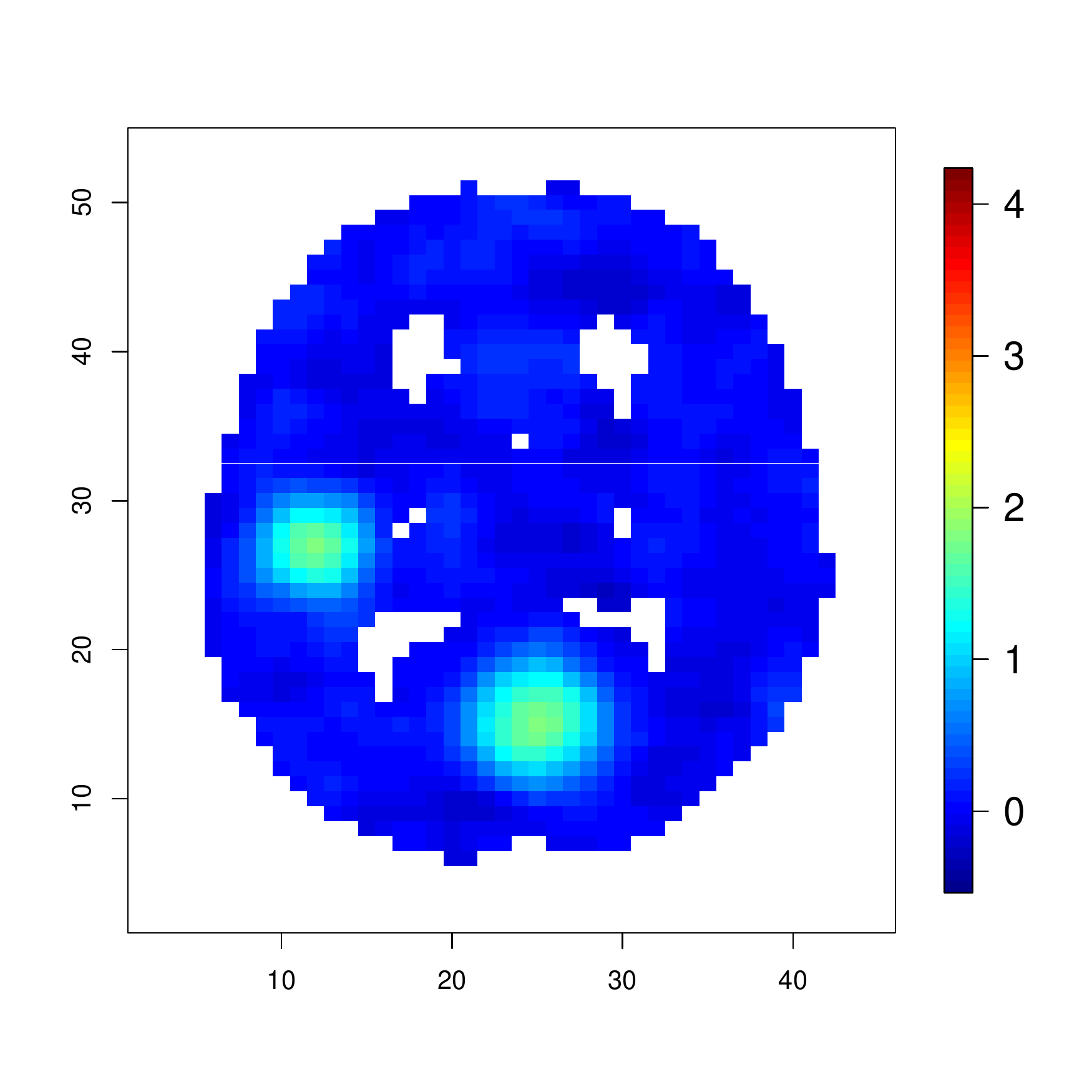} & 
\includegraphics[height=2in, trim=2.06cm 2.58cm 0.9cm 2cm, clip]{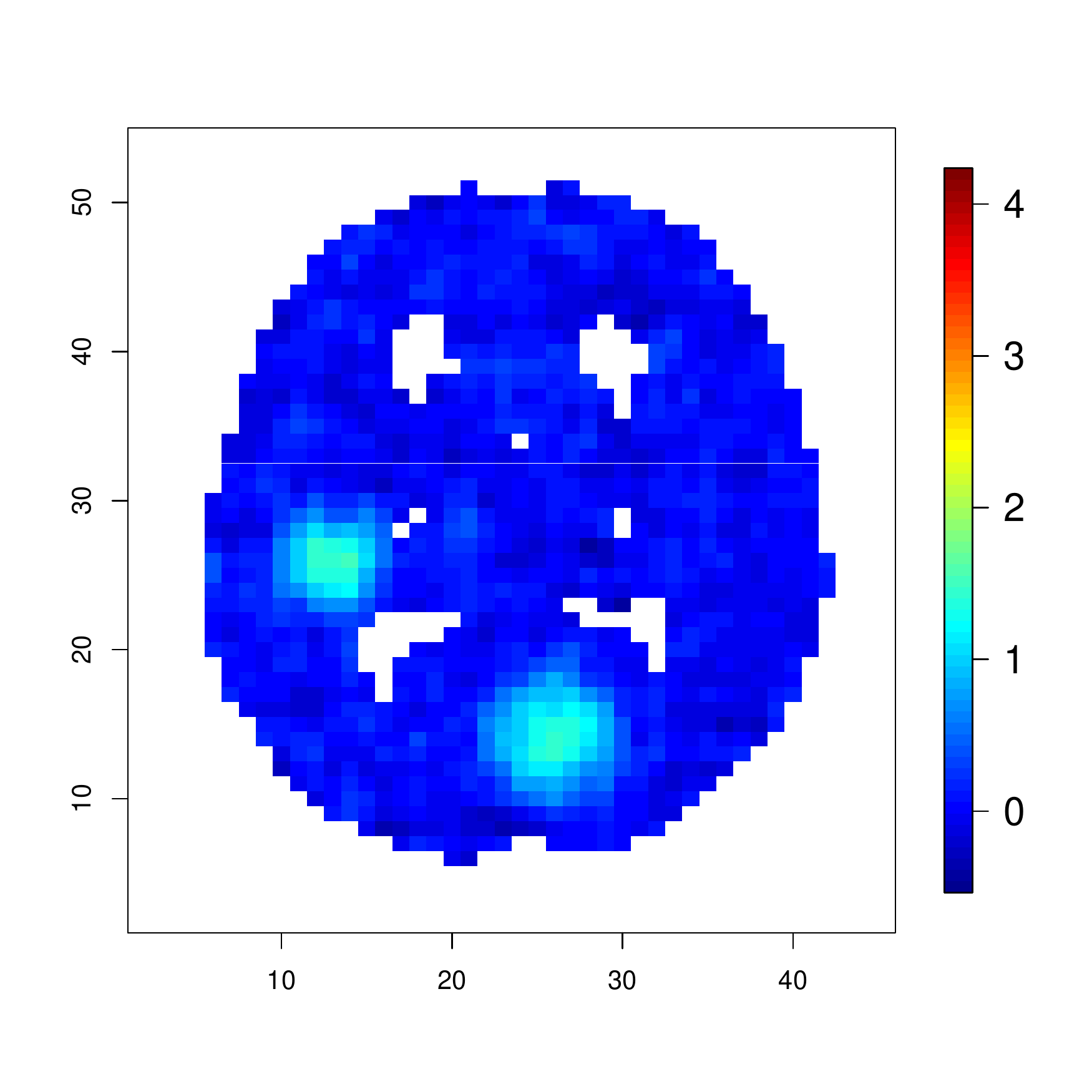} \\[-12pt]
\end{tabular}
\end{center}
\singlespace\caption{\small Bayesian and classical estimates of activation amplitudes in simulation study.}
\label{fig:sim:estimates}
\end{figure}

\begin{figure}
\begin{center}
\begin{tabular}{cc}
Activation 1\hspace{-7mm} & Activation 2\hspace{-7mm} \\[-6pt]
\includegraphics[width=3in]{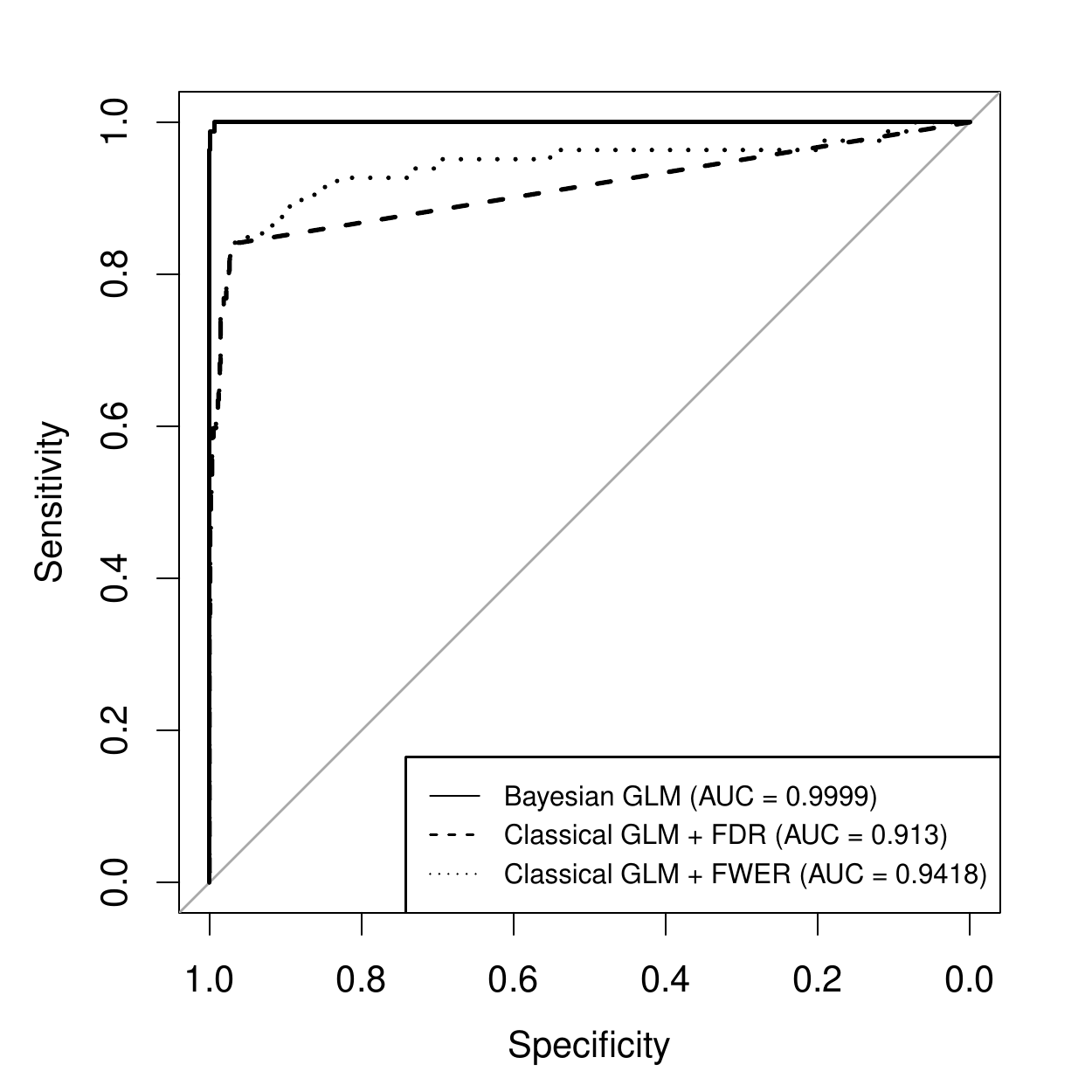} &
\includegraphics[width=3in]{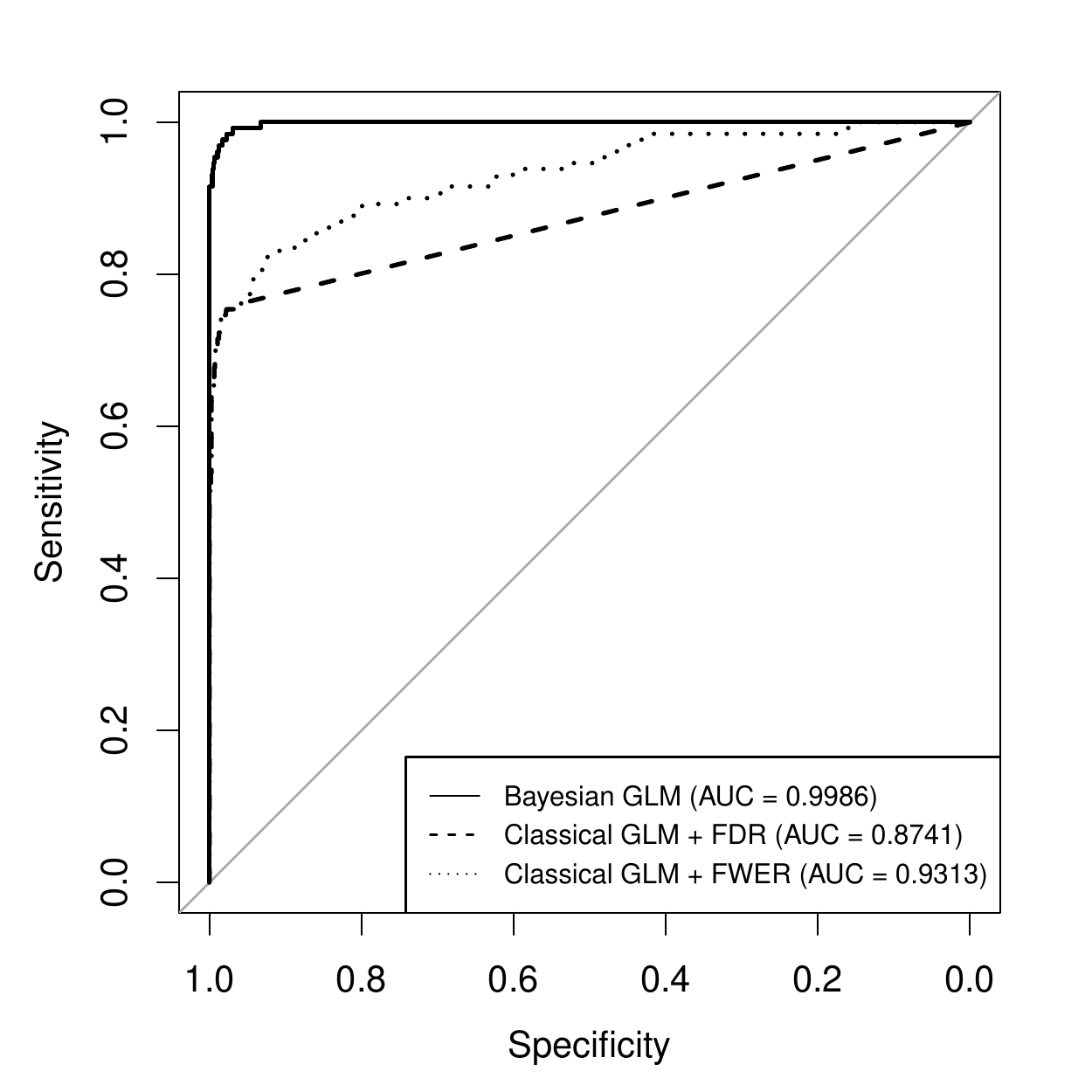}\\
\includegraphics[page=1, width=3in, trim=0 0 0 1cm, clip]{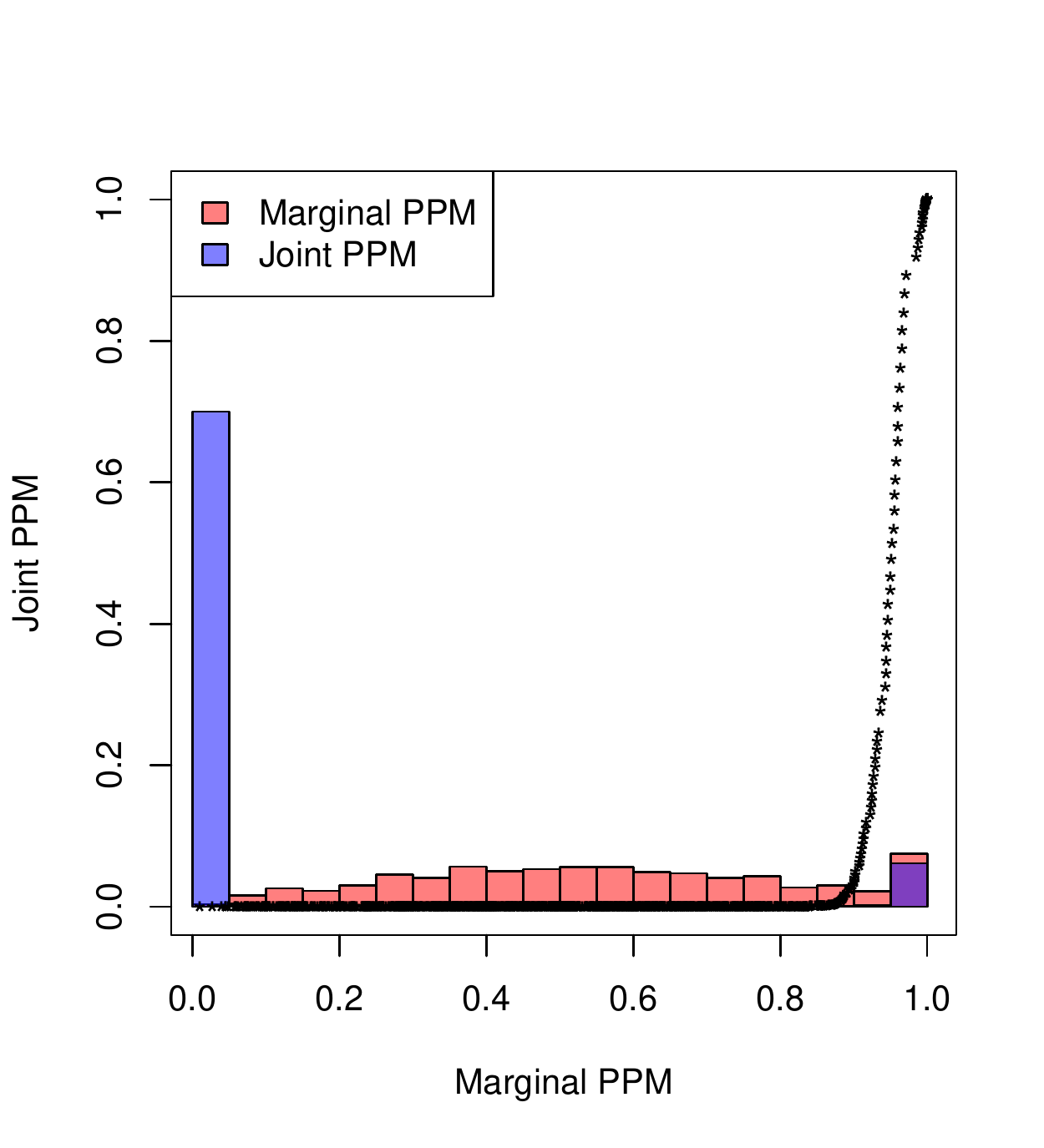} &
\includegraphics[page=2, width=3in, trim=0 0 0 1cm, clip]{reg_vs_jointPPM.pdf} \\[-20pt]
\end{tabular}
\end{center}
\singlespace\caption{\small {\bf Top row:} ROC curves for both activations of the simulation study.  In the Bayesian GLM, regions of activation are estimated by thresholding the excursion function at a certain level; in the classical GLM, the threshold is determined by performing a hypothesis test at each location, controlling the FDR or FWER of the image at a fixed level.  The marginal and joint PPMs have identical ROC curves, but as seen in the second row, the joint PPM provides more accurate false positive control and is less sensitive to the choice of threshold.  {\bf Bottom row:} Excursion function values obtained from marginal and joint PPM approaches.  The curves show that the excursion function based on the joint PPM is a monotonic transformation of that based on the marginal PPM, resulting in identical ROC curves.  However, the distribution of excursion function values obtained from the two approaches is very different.  The joint PPM (histogram shown in blue) results in values close to 0 and 1, and hence is less sensitive to the choice of significance threshold, while the marginal PPM (histogram shown in red) results in a more continuous distribution of values between 0 and 1. }
\label{fig:sim_roc}
\end{figure}

Estimates of the two amplitude fields based on the Bayesian and classical GLMs are shown in the top row of Figure \ref{fig:sim:estimates}. Both sets of estimates capture the general spatial patterns of the true fields, but the classical estimates are more noisy and also markedly underestimate the activation magnitudes.  The difference in performance of the two methods is more clearly illustrated in Figure \ref{fig:sim_roc}, which displays receiver operating characteristic (ROC) curves for the Bayesian and classical GLMs, based on varying the threshold of the excursion function and corrected $p$-value maps, respectively.  These curves illustrate that the Bayesian approach achieves both high sensitivity and specificity, with over $0.998$ area under the curve (AUC) for both activations.  In contrast, the classical GLM is not able to simultaneously achieve high sensitivity and high specificity, with AUC for each activation of $0.913$ and $0.874$ using FDR correction and $0.942$ and $0.931$ using FWER correction, respectively.  Furthermore, the difference in performance is greatest when specificity is close to $1$.  As high specificity is often a principal goal in fMRI task activation studies, this suggests that in practice, the proposed Bayesian approach will tend to achieve higher sensitivity than the classical GLM when the false positive rate, and hence specificity, is controlled at a fixed level.


The bottom row of Figure \ref{fig:sim_roc} compares the excursion function values obtained from the joint posterior probabilities and marginal posterior probabilities.  The points show that the excursion function based on the joint posterior probabilities is a monotonic transformation of the excursion function based on the marginal posterior probabilities, leading to identical ROC curves for both approaches.  This is due to the choice of parametric family implemented in the \verb@excursions@ R package, which has the monotonic property \citep{bolin2015excursion}.  However, the distribution of values obtained from the two approaches is very different.  The excursion function values based on the joint PPM (histogram shown in blue) are clustered around 0 and 1, while the values based on the marginal PPM (histogram shown in red) are continuously distributed between 0 and 1.  Therefore, besides providing more accurate control of false positives, the joint PPM is also less sensitive to the choice of significance threshold.

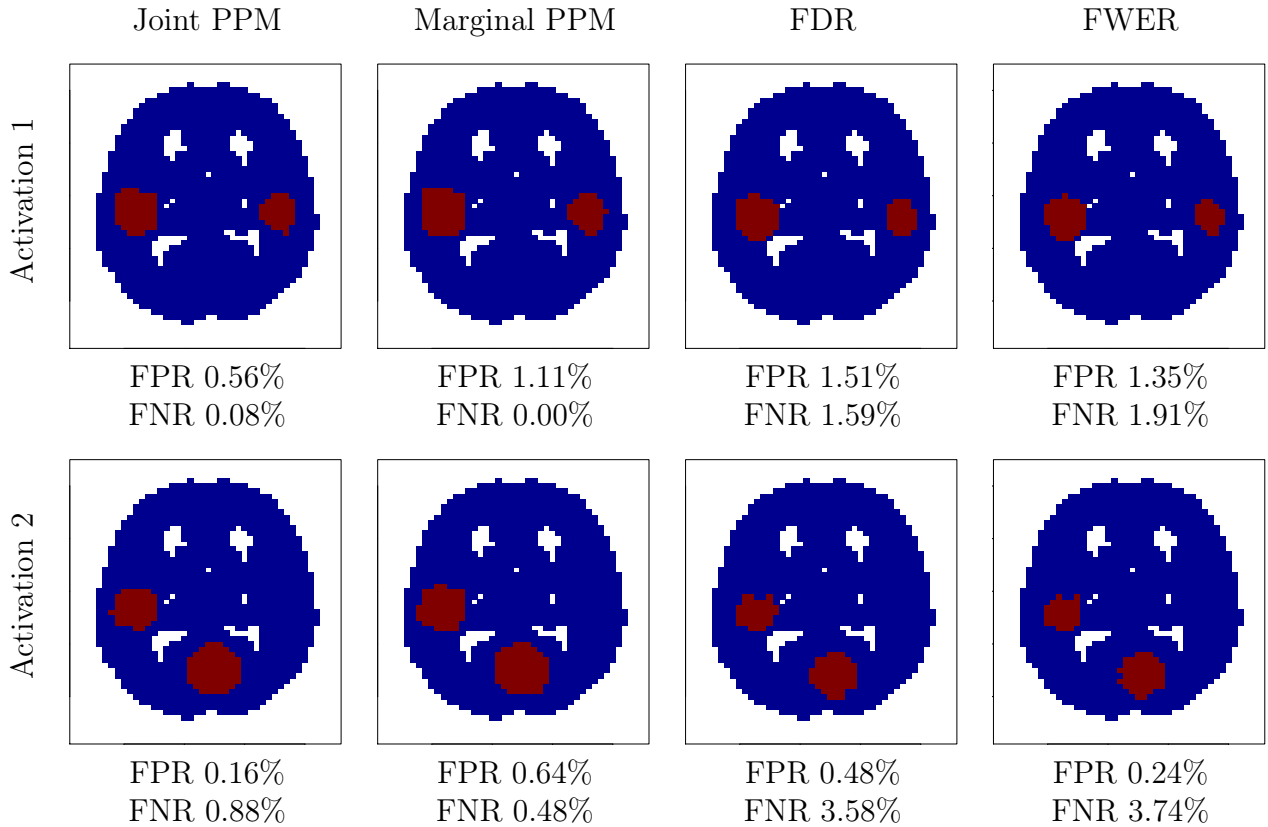
\begin{figure}
\begin{center}
\begin{tabular}{ccccc}
& Joint PPM & Marginal PPM & FDR & FWER \\[8pt]
\begin{picture}(10,105)
  \put(0,52){\rotatebox[origin=c]{90}{Activation 1}}
\end{picture} &
\includegraphics[height=1.5in, trim=2.06cm 2.58cm 3cm 2cm, clip]{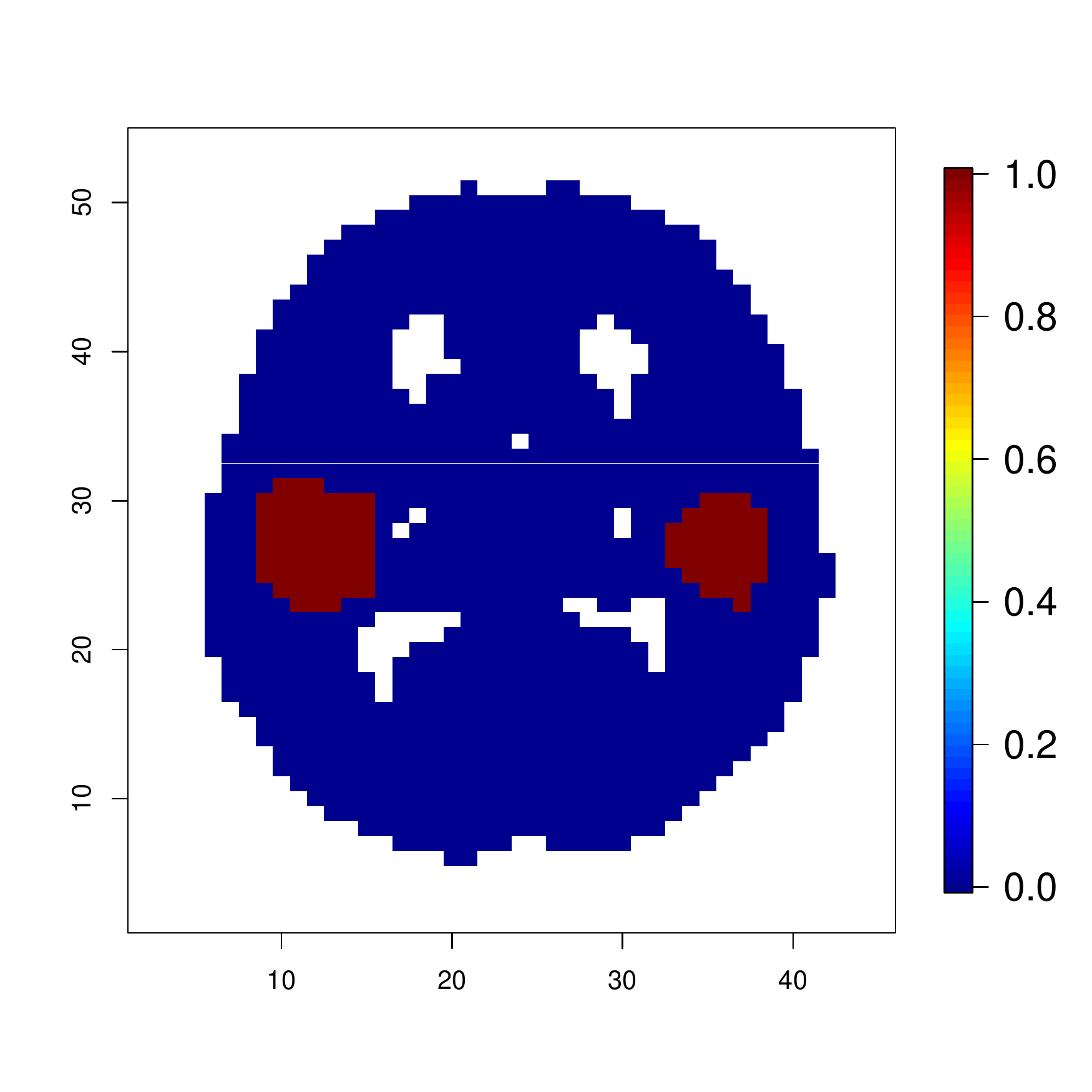}&
\includegraphics[height=1.5in, trim=2.06cm 2.58cm 3cm 2cm, clip]{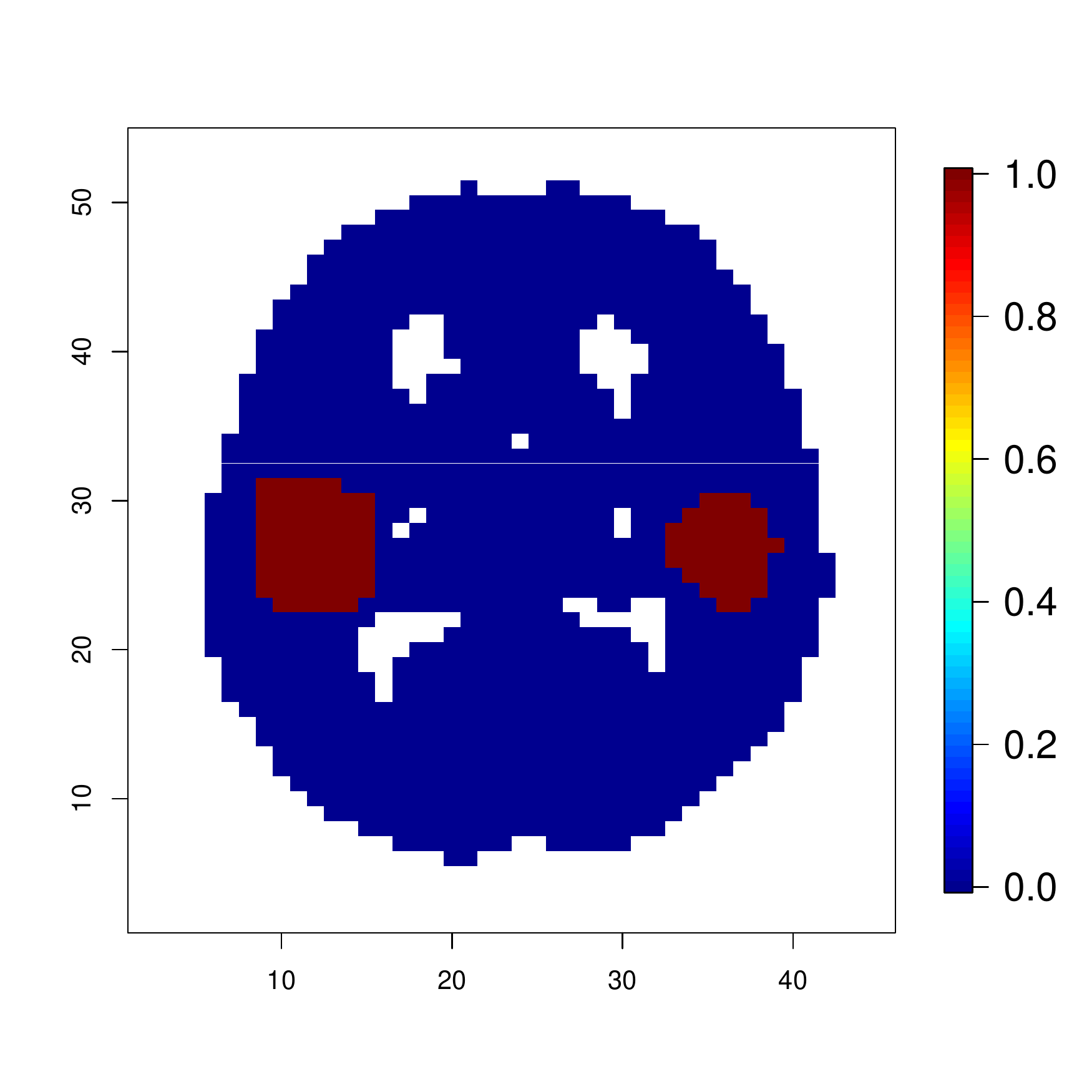}&
\includegraphics[height=1.5in, trim=2.06cm 2.58cm 3cm 2cm, clip]{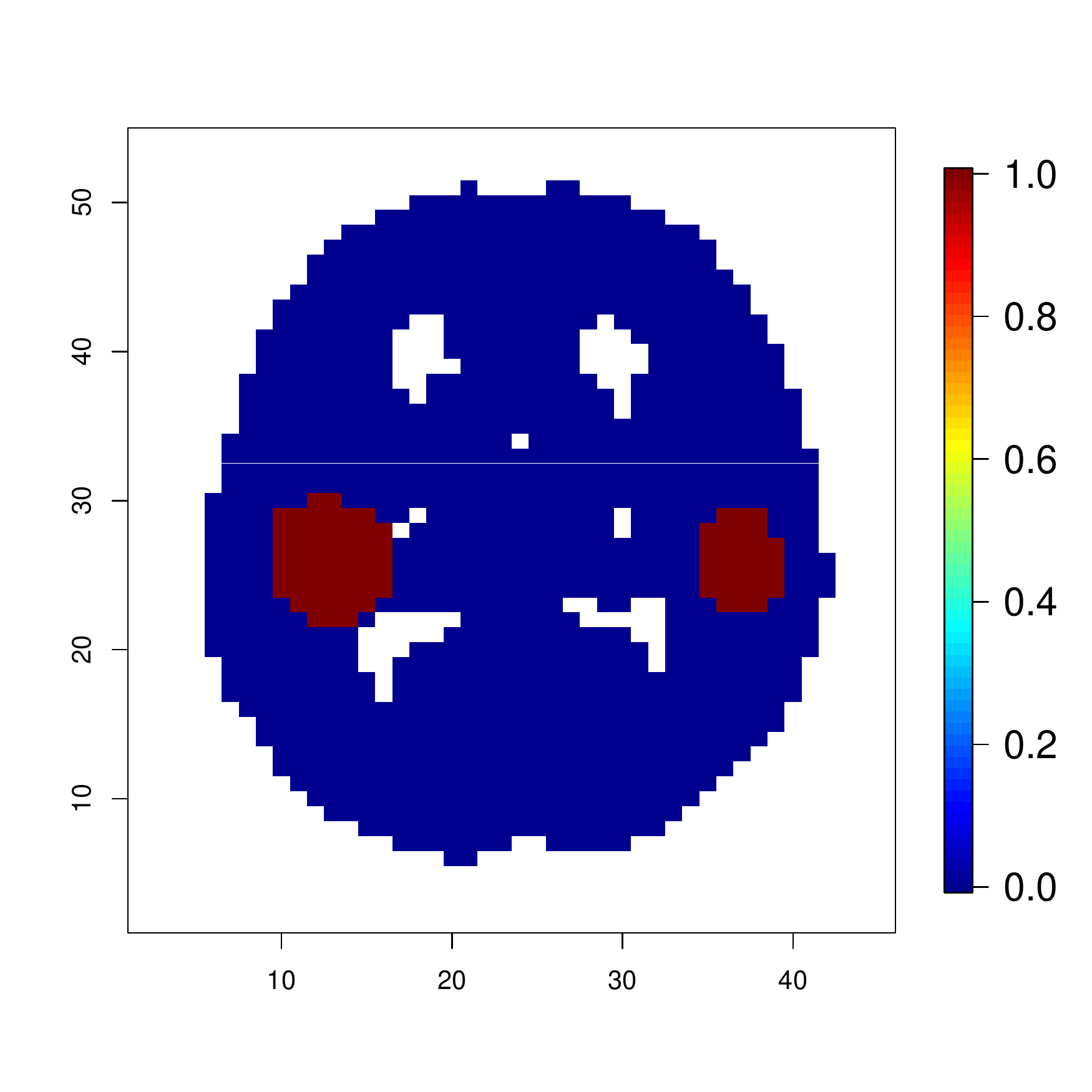} &
\includegraphics[height=1.5in, trim=2.06cm 2.58cm 3cm 2cm, clip]{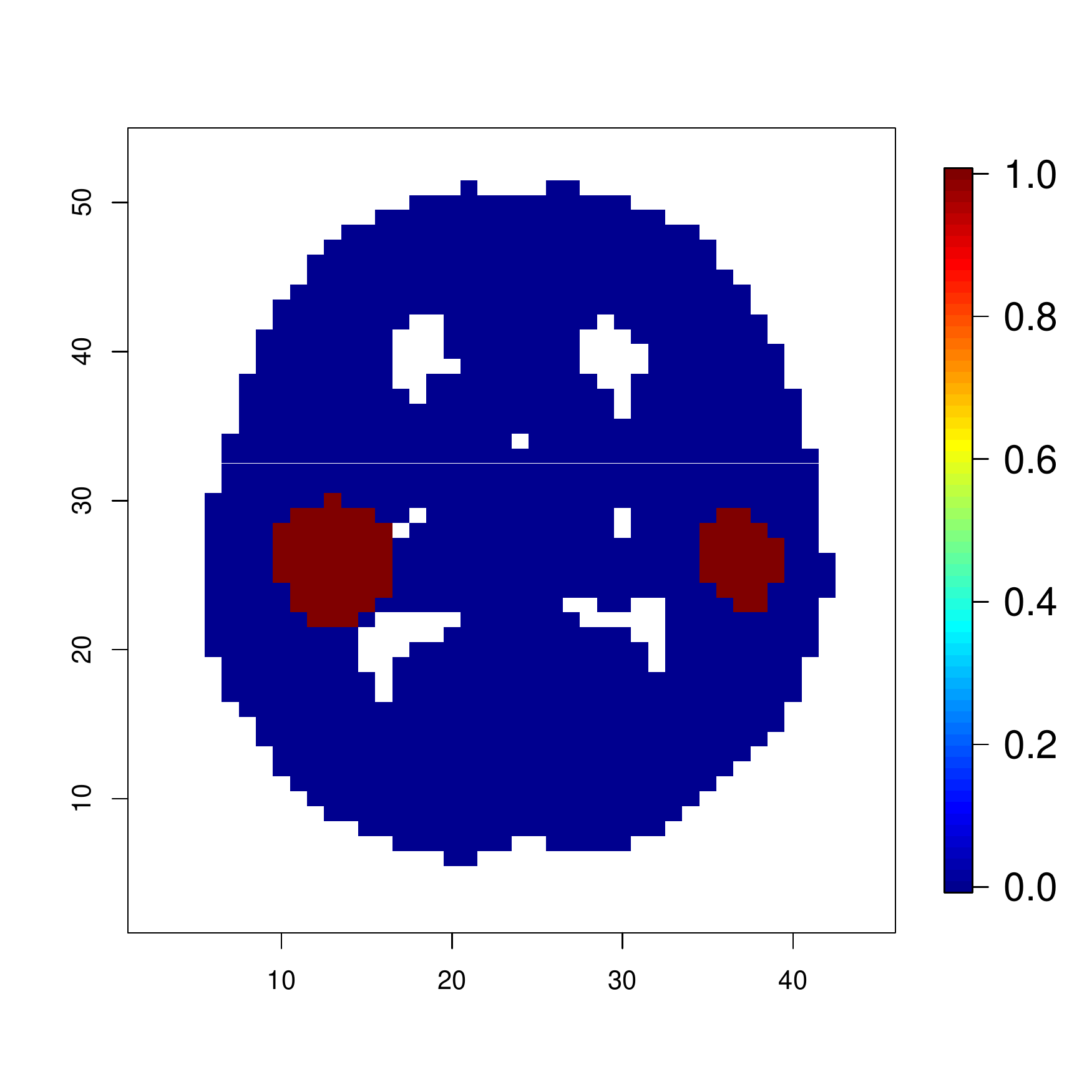} \\
& FPR 0.56\% & FPR 1.11\% & FPR 1.51\% & FPR 1.35\% \\
& FNR 0.08\% & FNR 0.00\% & FNR 1.59\% & FNR 1.91\% \\[8pt]
\begin{picture}(10,105)
  \put(0,52){\rotatebox[origin=c]{90}{Activation 2}}
\end{picture} &
\includegraphics[height=1.5in, trim=2.06cm 2.58cm 3cm 2cm, clip]{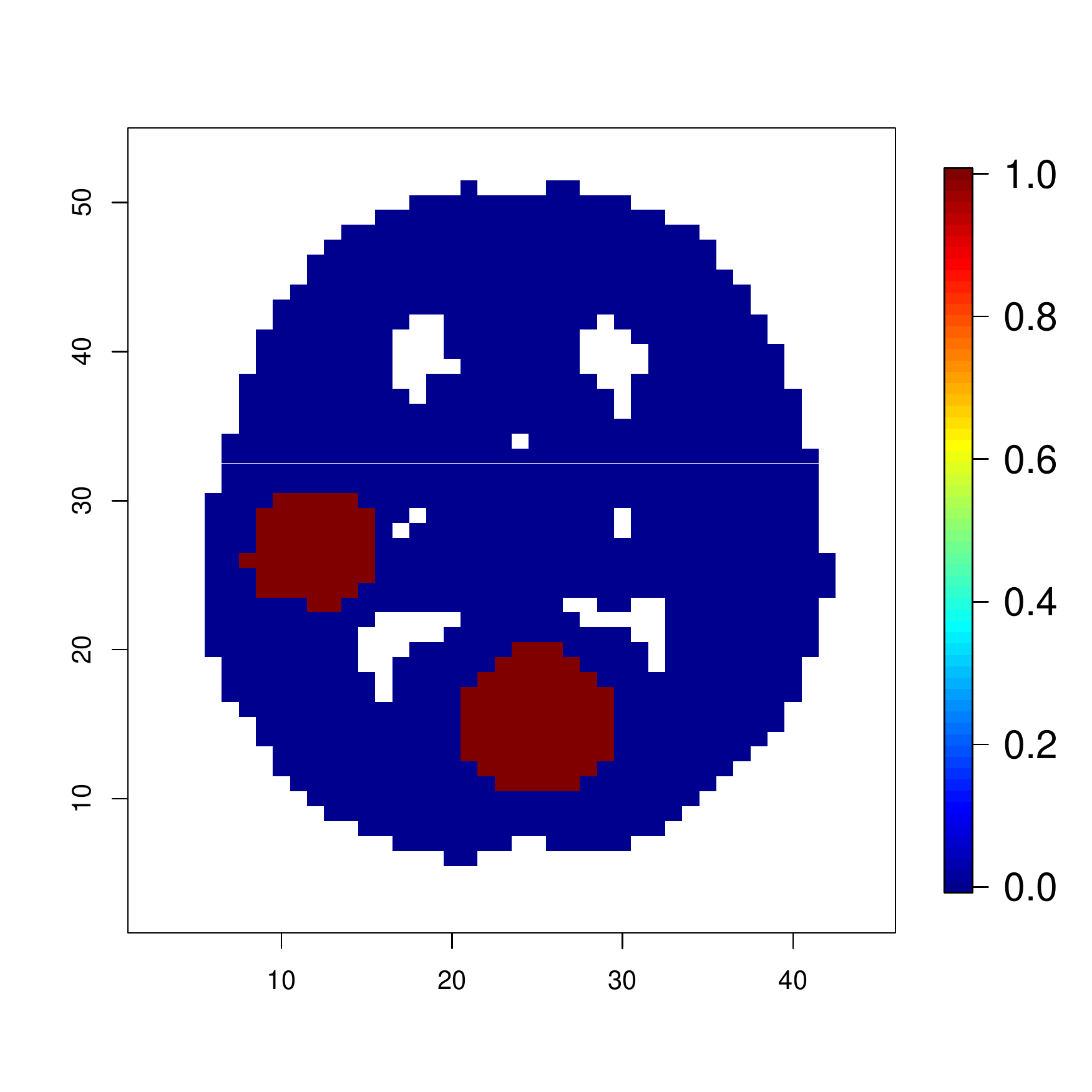}&
\includegraphics[height=1.5in, trim=2.06cm 2.58cm 3cm 2cm, clip]{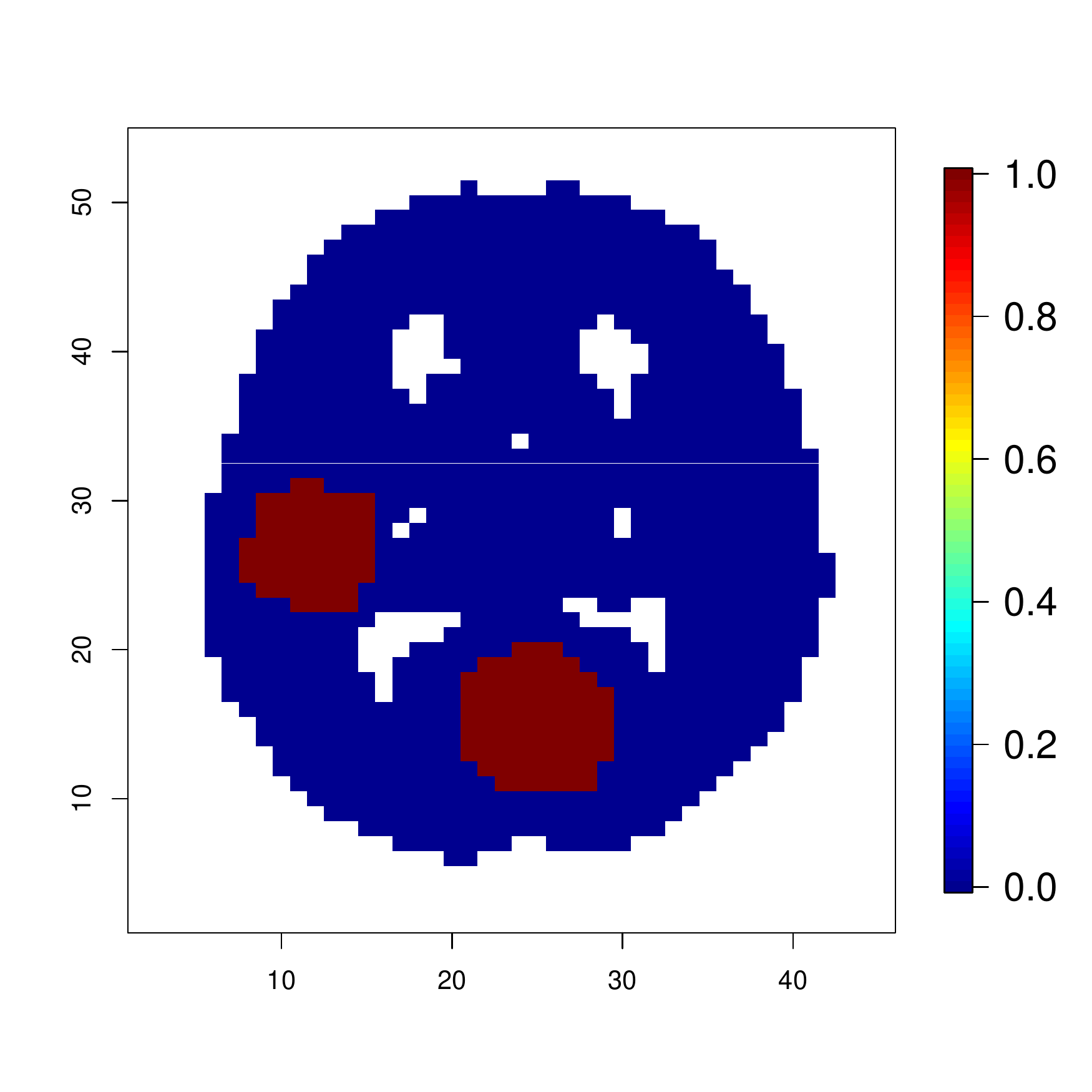}&
\includegraphics[height=1.5in, trim=2.06cm 2.58cm 3cm 2cm, clip]{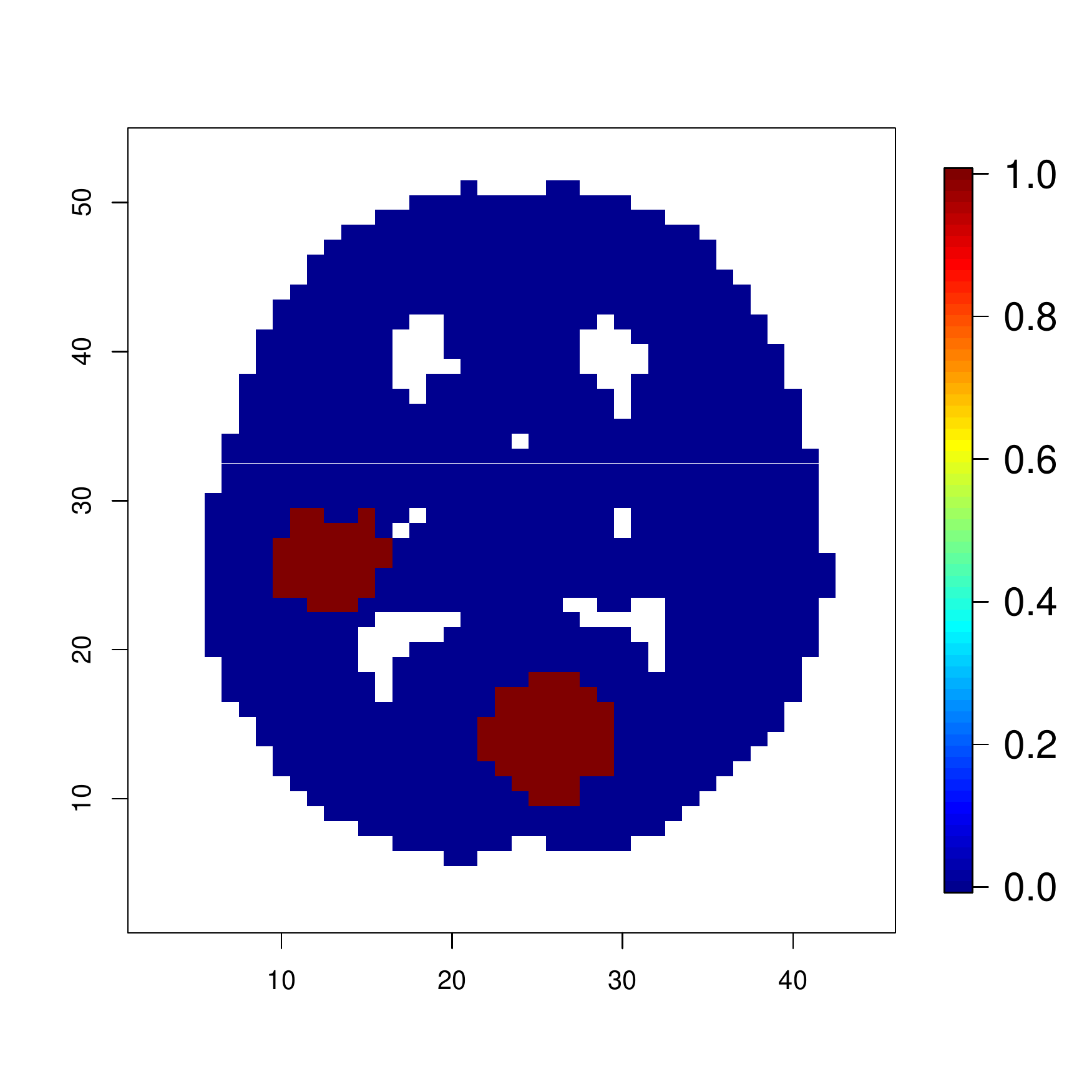} &
\includegraphics[height=1.5in, trim=2.06cm 2.58cm 3cm 2cm, clip]{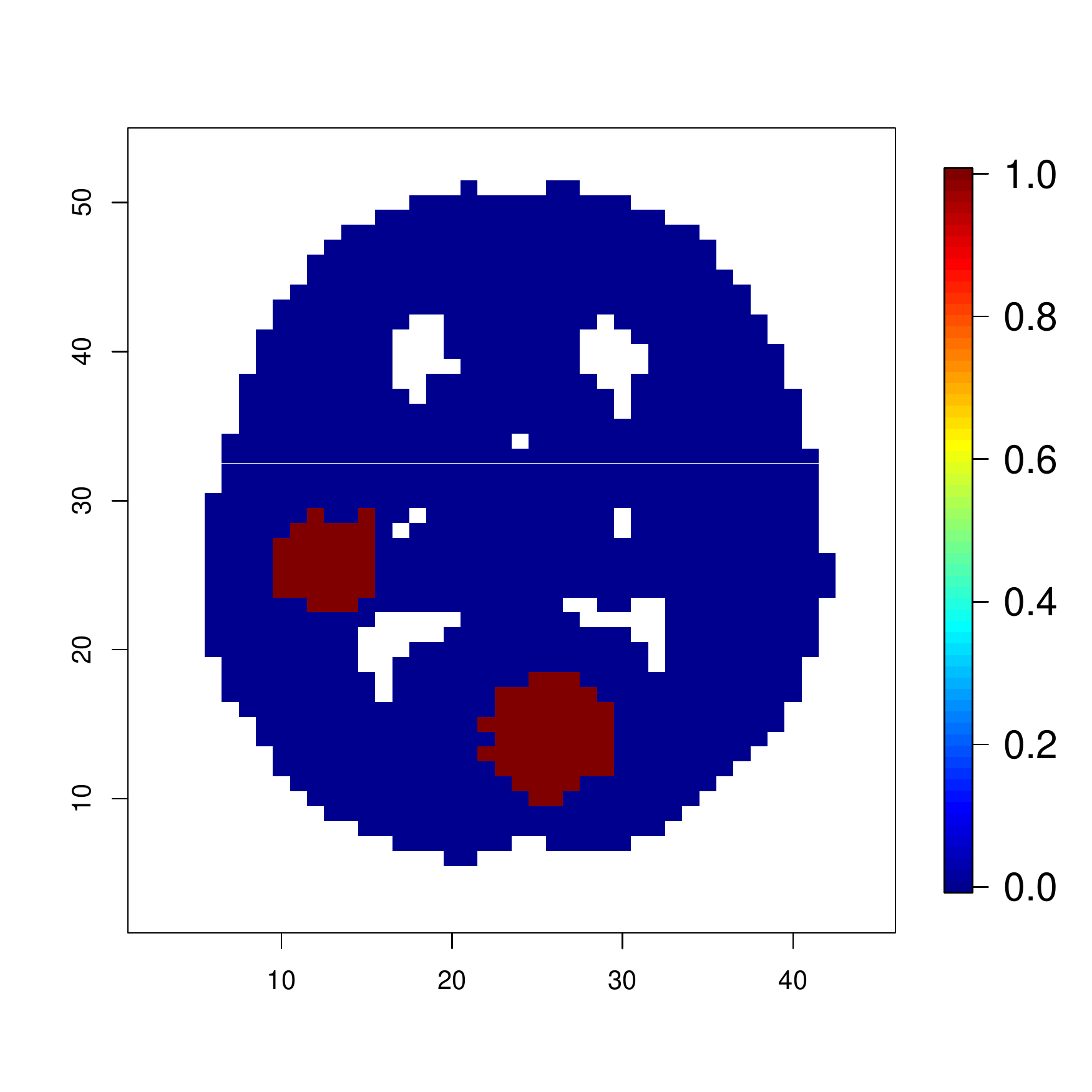} \\
& FPR 0.16\% & FPR 0.64\% & FPR 0.48\% & FPR 0.24\% \\
& FNR 0.88\% & FNR 0.48\% & FNR 3.58\% & FNR 3.74\%\\[-12pt]
\end{tabular}
\end{center}
\singlespace \caption{\small Estimated regions of activation in the simulation study.  In the Bayesian GLM, regions of activation are estimated using the joint and marginal PPM approaches with $\alpha=0.01$.  In the classical GLM, regions of activation are estimated by performing a hypothesis test on the task coefficient at each location, correcting for multiple comparisons through FDR control ($q=0.01$) and FWER control ($\alpha=0.01$).  The false positive rate (FPR) and false negative rate (FNR) are reported below each estimate, relative to the true activated regions shown in Figure \ref{fig:act}.}
\label{fig:sim99}
\end{figure}

Figure \ref{fig:sim99} displays estimated regions of activation obtained by thresholding the joint PPMs, marginal PPMs, and corrected $p$-values using a significance level of $0.01$ (see Figure S2 in Appendix B for regions of activation based on significance level $0.05$).  The false positive rate (FPR) and false negative rate (FNR) are reported for each method and activation, relative to the true active regions shown in Figure \ref{fig:act}.  The Bayesian joint PPM method significantly outperforms the classical GLM using either FDR or FWER correction, simultaneously achieving lower FPR and FNR for both activations.  This illustrates that the classical GLM approach not only results in a higher rate of false positives, but also suffers from reduced power.  Finally, as expected, the marginal PPM approach yields more false positives than the joint PPM approach because it fails to appropriately correct for multiple comparisons.   


\section{Experimental Data Results}\label{sec:app}

\subsection{Data Collection and Processing}\label{subsec:data}

We applied the proposed methods to fMRI data from $20$ randomly sampled subjects from the Human Connectome Project (HCP), a database of demographic, behavioral and neuroimaging data collected on over one thousand healthy adult subjects (\url{http://humanconnectome.org}).  We employed the fully preprocessed cortical surface fMRI data available in the HCP 500-subject data release.  A customized Siemens Skyra 3T scanner using a multi-band/multi-slice pulse sequence with an acceleration factor of eight was used to acquire fMRI scans with high temporal and spatial resolution \citep{moeller2010multiband, feinberg2010multiplexed, setsompop2012blipped, xu2012highly, uugurbil2013pushing}.  High-resolution structural $T_1$-weighted images were acquired for anatomical localization and transformation to standard space.  

The data first underwent the HCP \textit{fMRISurface} minimal preprocessing pipeline, which includes removal of spatial distortions, motion realignment, distortion correction, alignment to the structural image, bias field correction, intensity normalization and removal of extra-cerebral voxels \citep{glasser2013minimal}.  These steps are standard practice in fMRI data analysis and are necessary to align the data into a common space and remove major sources of noise.  
The volumetric fMRI data was then mapped from the volumetric representation to the cortical surface manifold, and subjects registered to the standard template space, as illustrated in Figure \ref{fig:vol_to_surf}.  To regularize the mapping process, a small degree of spatial smoothing was applied after transformation using a geodesic Gaussian surface smoothing algorithm with 2mm FWHM \citep{glasser2013minimal}.  While the spherical representation is used for inter-subject registration and fitting the spatial Bayesian model, lesser degrees of inflation, which maintain basic brain shape and structure, are used here for display purposes.

During each subject's $3.5$-minute fMRI run, a series of $284$ functional volumes were acquired, one every $0.72$ seconds.  Subjects performed a series of five motor tasks during the acquisition, each preceded by a three-second visual cue \citep{HCPTaskProtocol}. The following five tasks were performed twice over the run for twelve seconds each time: move tongue, tap left fingers, tap right fingers, wiggle left toes, and wiggle right toes. For each type of task and the visual cue, a binary stimulus function was defined taking the value $1$ during the periods during which the subject was asked to perform the task or view the cue and $0$ otherwise.  These six binary stimulus functions were each convolved with a canonical hemodynamic response function \citep{friston1998event}, resulting in a set of stimulus response functions for each subject (see Figure S3 in Appendix C).  Three 15-second periods during which the subject was asked to fixate on a crosshair displayed on the screen served as a baseline.  

To account for noise due to subject motion, six rigid body realignment parameters estimated in the motion realignment phase of preprocessing and their first-order temporal derivatives were included in the model as nuisance covariates.  Linear and quadratic time terms were also included to account for scanner drift.  To reduce the computational cost of fitting the Bayesian model, the fMRI time courses were first pre-whitened by assuming an AR($p$) process on the residuals from a classical GLM with uncorrelated errors.  A model order of $p=6$ was chosen based on inspection of the partial autocorrelation function.  This relatively high model order is not surprising due to the fast temporal resolution of the data.  The AR coefficients were estimated by solving the Yule-Walker equations and were allowed to vary spatially, as previous studies have shown that the degree of temporal autocorrelation is not constant across the brain \citep{worsley2002general, eklund2012does}.  To improve estimation efficiency, the AR coefficient estimates at each location were averaged across all subjects in the analysis.  

\subsection{Model Estimation and Results}

\begin{figure}
\begin{center}
\begin{tabular}{cc}
Left Hemisphere & Right Hemisphere \\[14pt]
\includegraphics[width=2in, trim=2in 2in 2in 2in, clip]{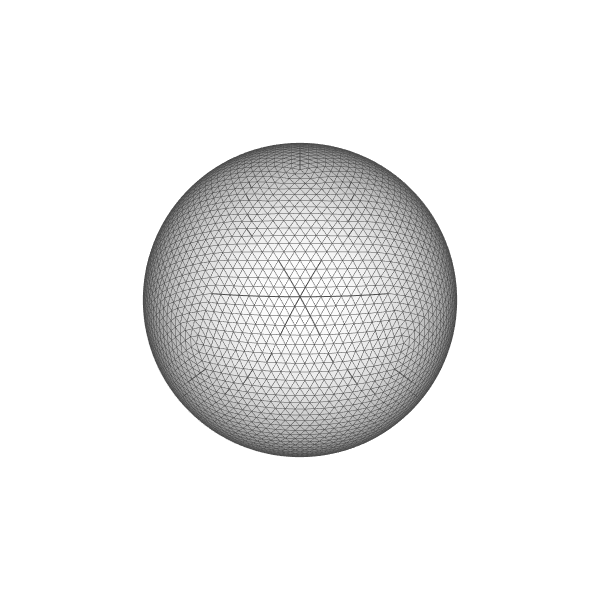} \hspace{5mm} & \hspace{5mm}
\includegraphics[width=2in, trim=2in 2in 2in 2in, clip]{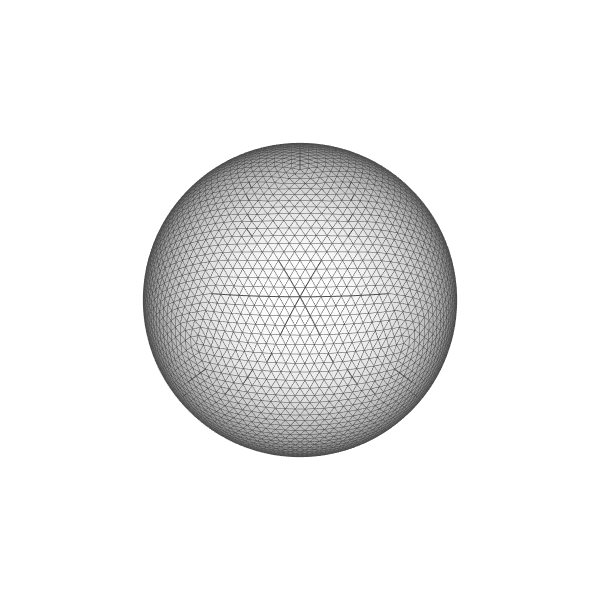}\\[14pt]
\includegraphics[width=2in, trim=2in 2in 2in 2in, clip]{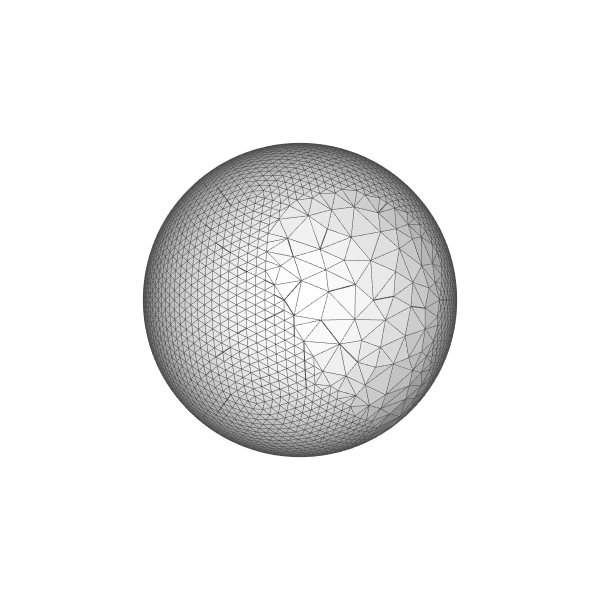} \hspace{5mm} & \hspace{5mm}
\includegraphics[width=2in, trim=2in 2in 2in 2in, clip]{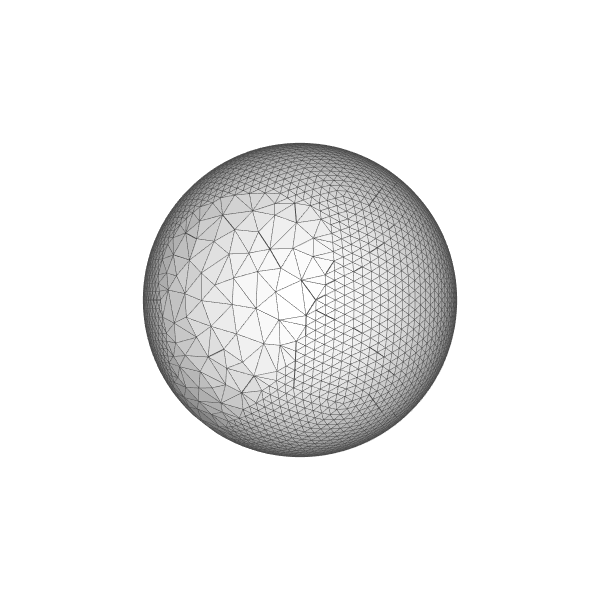}
\end{tabular}
\end{center}
\singlespace\caption{\small The spherical mesh for each hemisphere. For each hemisphere, the lateral (exterior) cortical surface is displayed on top and the medial (interior) cortical surface is displayed on bottom.}
\label{fig:app:mesh}
\end{figure}

The left and right hemisphere each contain 32,492 surface vertices, including approximately 2,700 missing values in the medial wall.  To reduce the computational burden while maintaining high spatial resolution, we used the Connectome Workbench to resample each hemisphere to approximately 6,000 vertices \citep{marcus2011informatics}.  A spherical mesh was created based on the radial coordinates of the locations excluding missing values in the medial wall, resulting in a mesh with approximately 5,300 vertices.  Figure \ref{fig:app:mesh} displays the lateral (external) and medial (internal) surfaces of the spherical mesh for each hemisphere.  After model fitting, coefficient estimates and posterior probabilities for the 32,492 vertices for each hemisphere were obtained by projecting back to the original resolution using the Connectome Workbench tools.  


For the $m^{\text{th}}$ ($m=1,\ldots,20$) subject, we considered the following model for each hemisphere:
\be\label{mod:app}
\bfy_m = \sum_{k=0}^6\bfX_{mk}\bfbeta_{mk} + \sum_{j=1}^{14}\bfZ_{mj}\bfb_{mj} + \bfvareps_m,\quad\bfvareps_m\sim N(\bfzero, \sigma^2_m\bfI),
\ee
where $\bfy_m$ is a vector containing the $NT$ response values, where $T=284$ is the number of time points and $N\approx5300$ is the number of locations in the mesh; the $\bfX_{mk}$ are the stimulus response functions corresponding to baseline, the visual cue, and the five motor tasks; and the $\bfZ_{mj}$ contain the nuisance covariates for motion and drift.  Note that here uncorrelated errors are assumed because temporal autocorrelation was removed prior to model fitting through prewhitening. The SPDE spatial priors and diffuse Gaussian priors described in Section \ref{sec:method} were independently assumed on $\bfbeta_{mk}$ and $\bfb_{mj}$, respectively; other priors were specified as described in Section \ref{sec:method}. We used the INLA method to fit model (\ref{mod:app}) and obtain posterior estimates for each subject. Using the joint and two-level models described in Section \ref{sec:fmri:pop}, we then combined those subject-level estimates to derive group-level posterior estimates. Finally, we used the excursion set method based on the joint PPM described in Section \ref{sec:ppm} to identify group-level regions of activation in response to each task. Two values of the activation threshold $\gamma$ were considered: $0$ and 1\% of the baseline signal, similar to \cite{qui:ni:10}.

For comparison purposes, we also applied the classical GLM method to the data.  For each subject, we first spatially smoothed the data using the Connectome Workbench geodesic Gaussian surface smoothing algorithm with 6mm FWHM \citep{glasser2013minimal}.  We then prewhitened the data as described above.  We fit a classical GLM to each subject's data including the 14 nuisance terms as in the Bayesian GLM.  For each task, we then used the subject-level estimates in a group-level model with weights inversely proportional to the squared standard error of the subject-level coefficients.  Finally, we identified group-level regions of activation by performing a $t$-test at every vertex, accounting for multiple comparisons with FDR correction or FWER correction similarly as in the simulation described in Section \ref{sec:simu}.  To control the FWER at the group level, we performed a permutation test by generating $100$ resamplings of each subject's prewhitened data, refitting the subject-level models for each resampling, then re-estimating the group-level model to obtain the null distribution of the maximum $t$ statistic in the image.

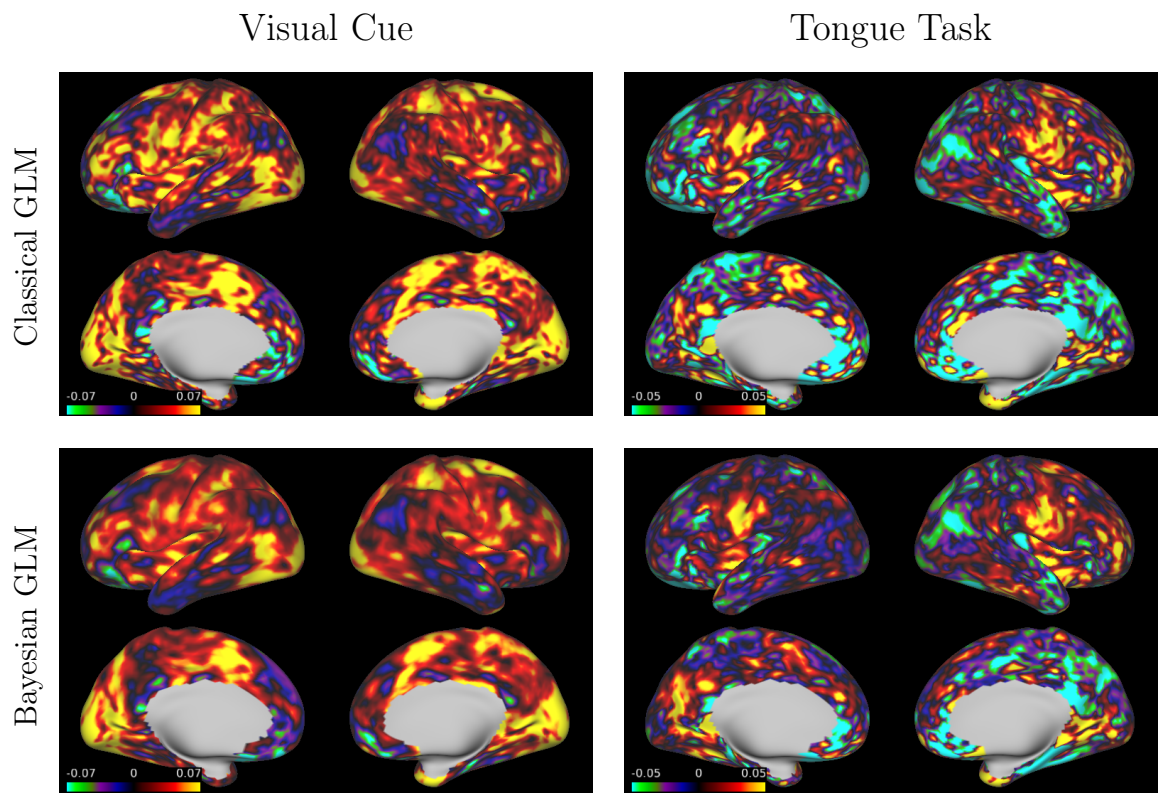
\begin{figure}
\begin{center}
\begin{tabular}{cccc}
& \large Visual Cue & \large Tongue Task\\[8pt]
\begin{picture}(5,120)
  \put(0,60){\rotatebox[origin=c]{90}{Classical GLM}}
\end{picture} &
\includegraphics[height=1.8in]{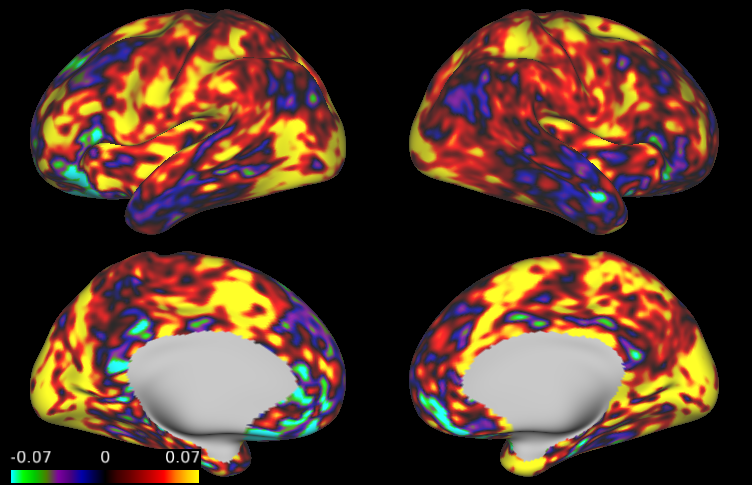}&
\includegraphics[height=1.8in]{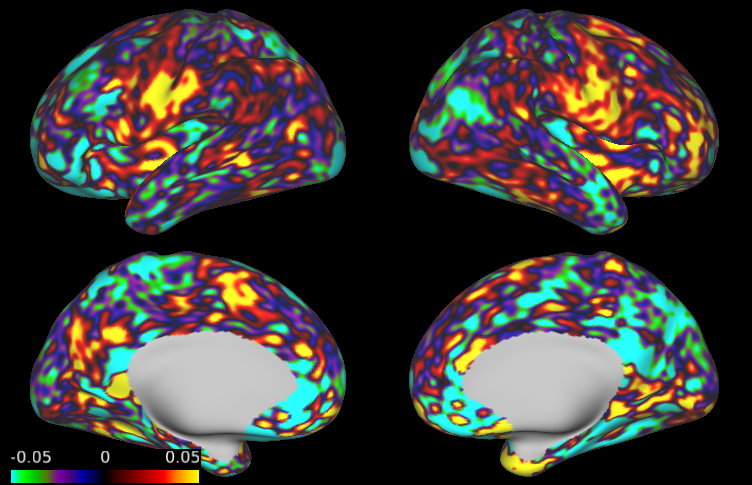}\\[8pt]
\begin{picture}(5,120)
  \put(0,60){\rotatebox[origin=c]{90}{Bayesian GLM}}
\end{picture} &
\includegraphics[height=1.8in]{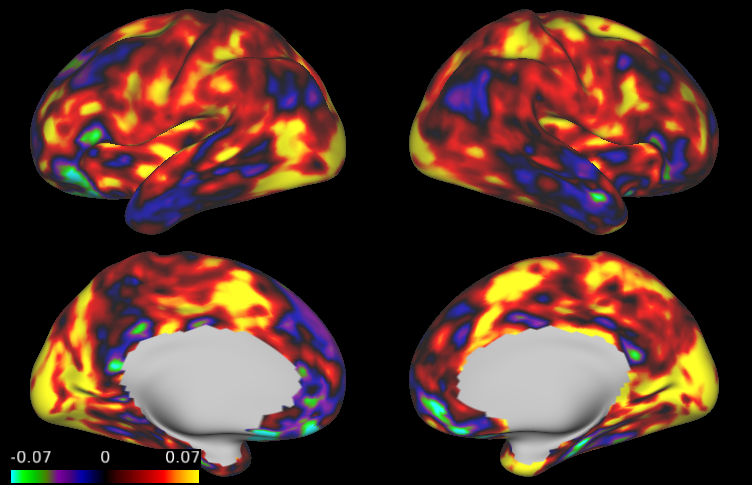}&
\includegraphics[height=1.8in]{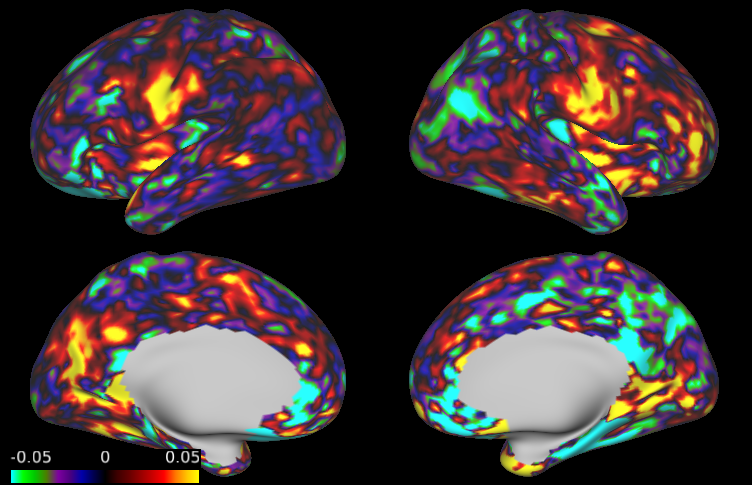}\\
\end{tabular}
\end{center}
\singlespace\caption{\small For one randomly selected subject, estimates of activation amplitude for each task, based on the classical and Bayesian GLM.}
\label{fig:app:beta_subj}
\end{figure}
\begin{figure}
\begin{center}
\begin{tabular}{cccc}
& \large Visual Cue & \large Tongue Task\\[8pt]
\begin{picture}(5,120)
  \put(0,60){\rotatebox[origin=c]{90}{Classical GLM}}
\end{picture} &
\includegraphics[height=1.8in]{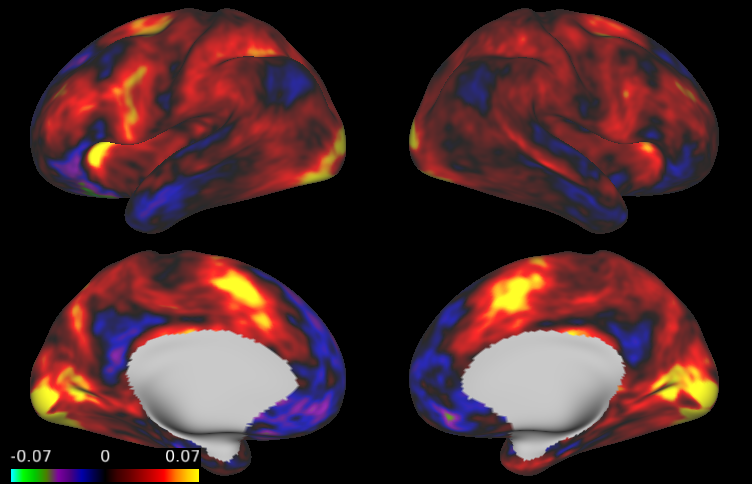}&
\includegraphics[height=1.8in]{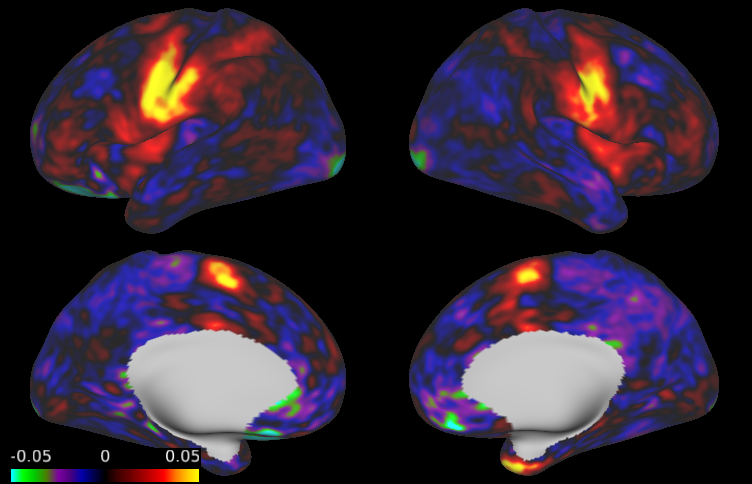}\\[8pt]
\begin{picture}(5,120)
  \put(-15,60){\rotatebox[origin=c]{90}{Bayesian GLM}}
  \put(0,60){\rotatebox[origin=c]{90}{(joint model)}}
\end{picture} &
\includegraphics[height=1.8in]{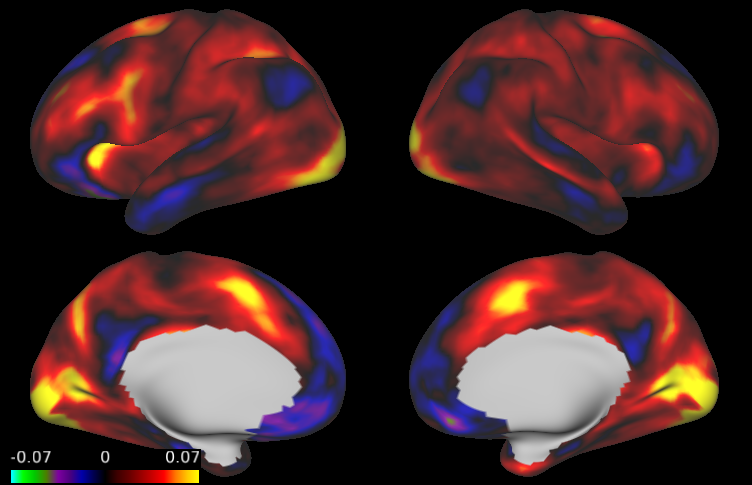}&
\includegraphics[height=1.8in]{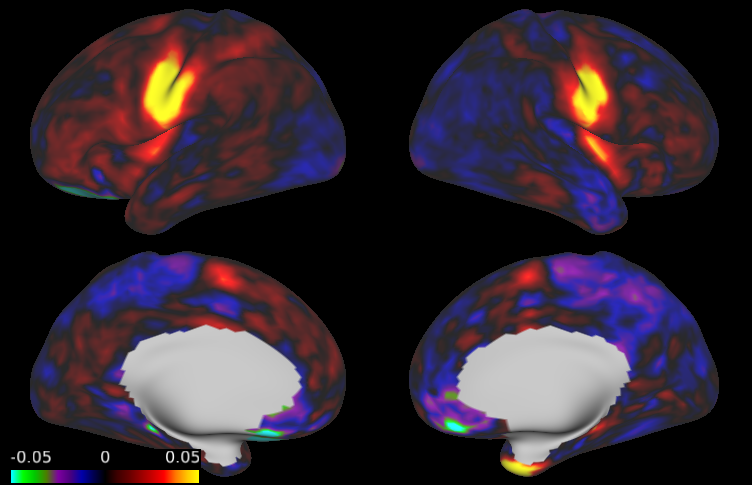}\\[8pt]
\begin{picture}(5,120)
  \put(-15,60){\rotatebox[origin=c]{90}{Bayesian GLM}}
  \put(0,60){\rotatebox[origin=c]{90}{(two-level model)}}
\end{picture} &
\includegraphics[height=1.8in]{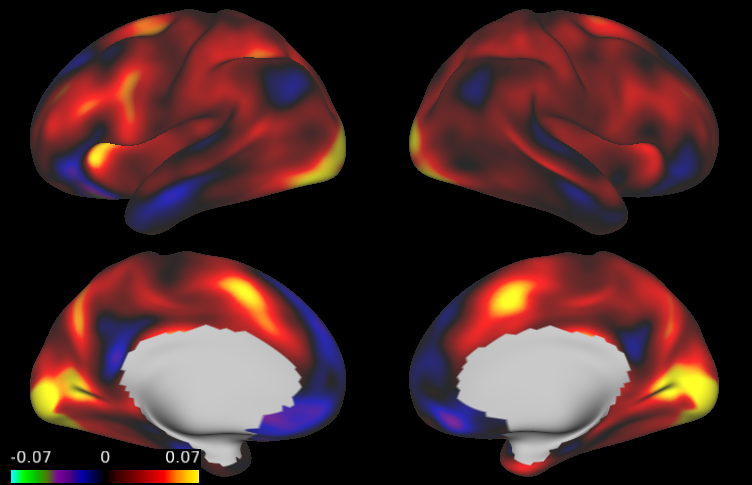}&
\includegraphics[height=1.8in]{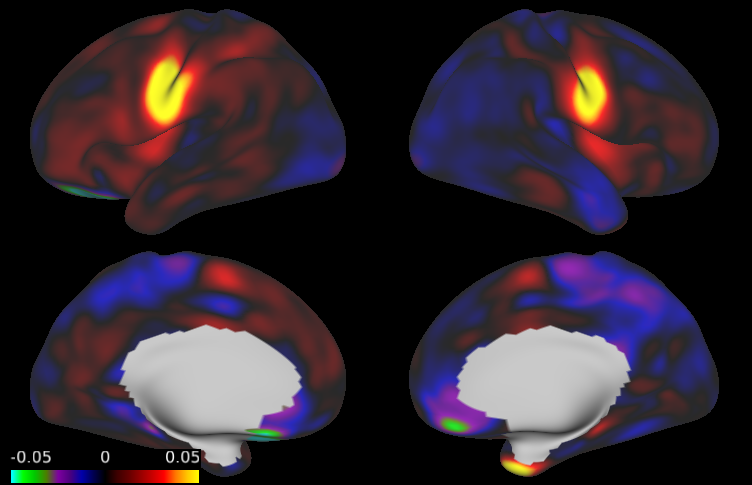}
\end{tabular}
\end{center}
\singlespace\caption{\small Group-level estimates of activation amplitude for each task, based on the classical and Bayesian GLM.}
\label{fig:app:beta}
\end{figure}
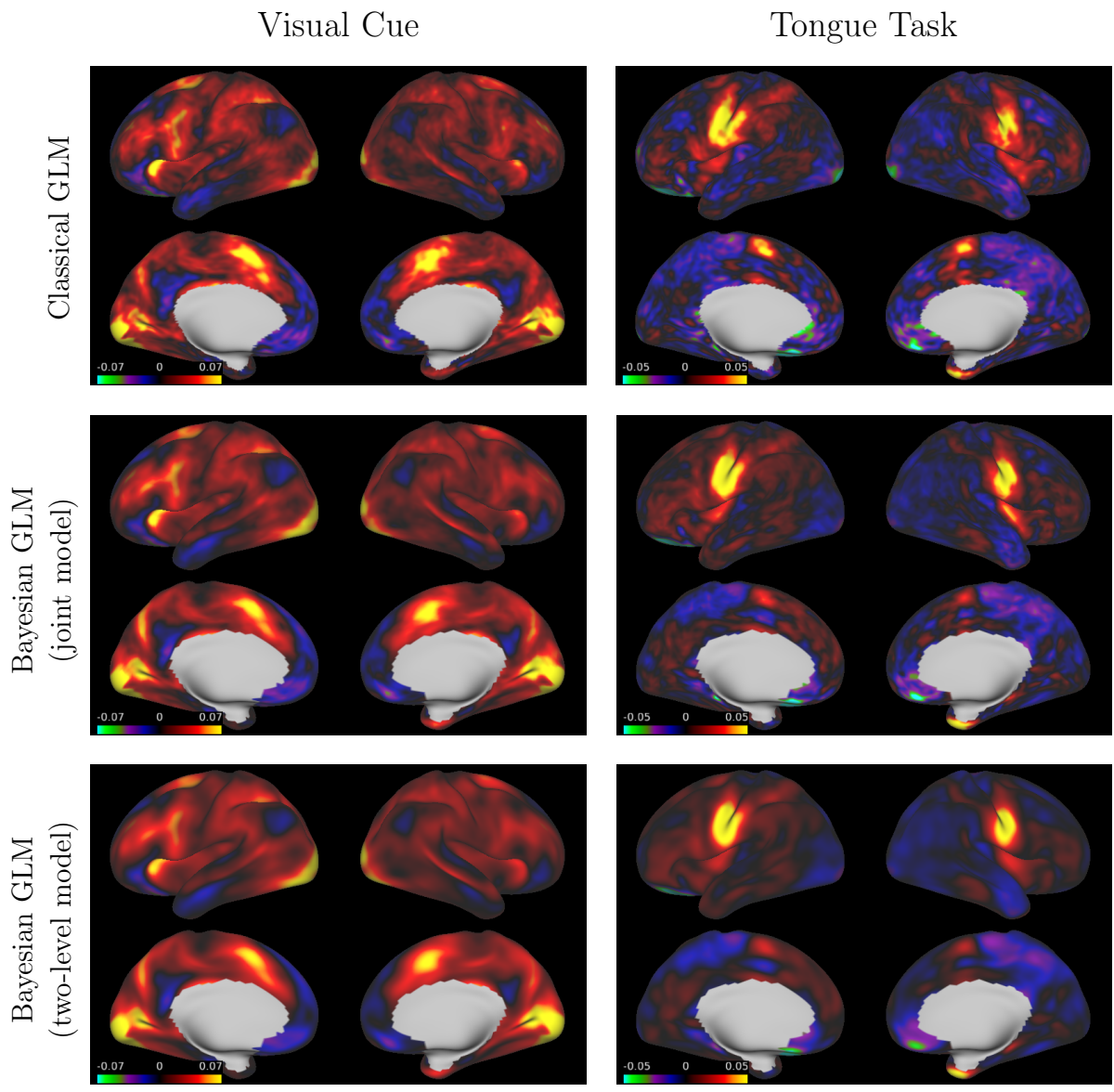

Figure \ref{fig:app:beta_subj} displays the Bayesian estimates of activation amplitude (posterior means) for the visual cue and tongue tasks, along with the classical GLM estimates, for one randomly selected subject.  The remaining tasks are shown in Appendix C.  Both sets of estimates are noisy, due to the high levels of noise in fMRI data and relatively short duration of the tasks performed for each individual subject ($30$ seconds for the visual cue; $24$ seconds for the tongue task); however, the Bayesian estimates are substantially smoother than the classical estimates, even though the data was smoothed prior to fitting the classical GLM but not the Bayesian GLM.  Figure \ref{fig:app:beta} displays group-level estimates of activation amplitude based on the classical and Bayesian GLM approaches.  For the Bayesian GLM, the results using the joint modeling approach and the two-level modeling approach with the sampling method described in Section \ref{sec:2level} are both displayed. In both the subject-level and group-level estimates, the activation fields for each task are generally as expected: during the visual cue, the visual and orbitofrontal (decision making) regions are highly active, while the somatomotor (motor planning) region is activated to a lesser degree; during the tongue movement task, the area of the motor cortex associated with the tongue shows a strong degree of activation; and the remaining motor tasks activate the expected areas of the motor cortex (see Figures S4 and S5 in Appendix C).  Compared with the classical GLM, the Bayesian GLM results in smoother estimates of activation, as it accounts for dependence in the degree of activation between neighboring locations.  As hypothesized, the two-level modeling approach tends to result in somewhat oversmoothed estimates, compared with the joint modeling approach.  


\begin{figure}
\begin{center}
\begin{tabular}{cccc}
& \large Visual Cue & \large Tongue Task\\[6pt]
\begin{picture}(5,120)
  \put(-15,60){\rotatebox[origin=c]{90}{Classical GLM}}
  \put(0,60){\rotatebox[origin=c]{90}{(FWER control)}}
\end{picture} &
\includegraphics[height=1.8in]{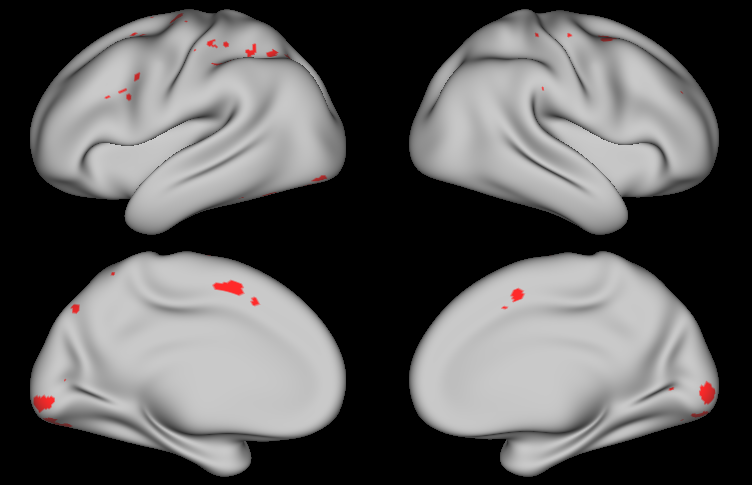}&
\includegraphics[height=1.8in]{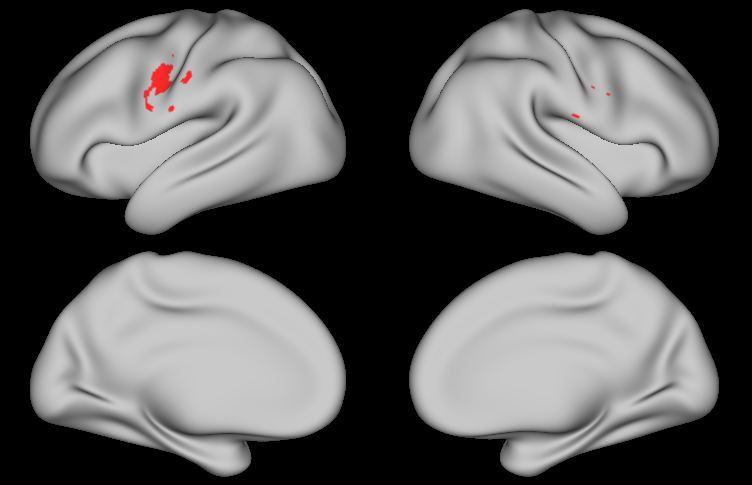}\\[4pt]
\begin{picture}(5,120)
  \put(-15,60){\rotatebox[origin=c]{90}{Classical GLM}}
  \put(0,60){\rotatebox[origin=c]{90}{(FDR control)}}
\end{picture} &
\includegraphics[height=1.8in]{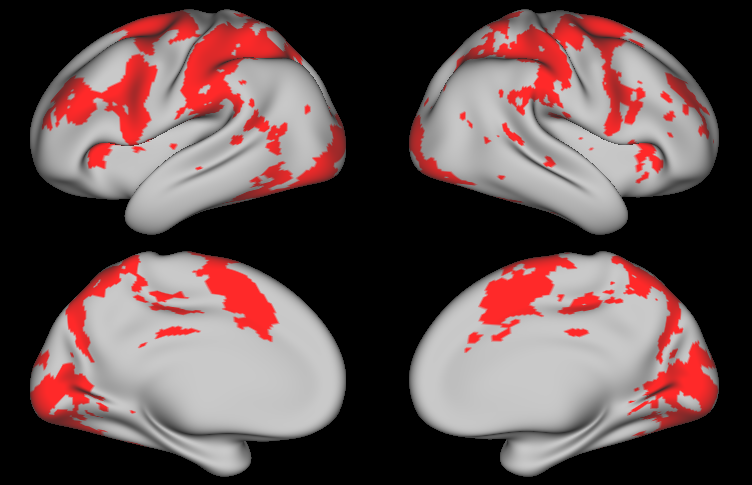}&
\includegraphics[height=1.8in]{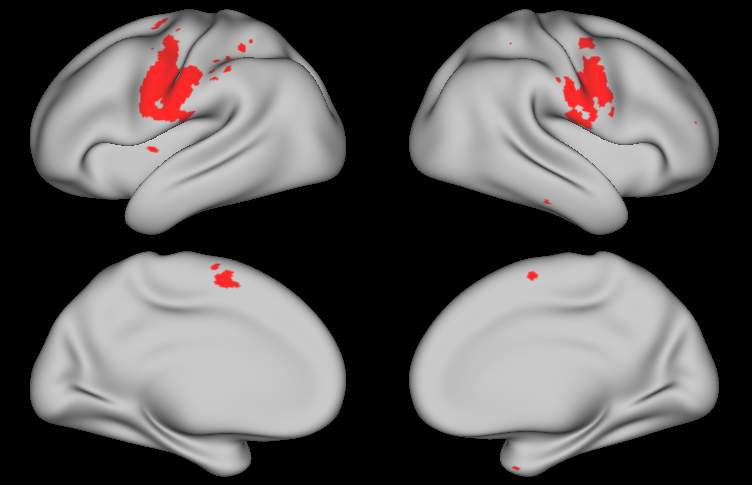}\\[4pt]
\begin{picture}(5,120)
  \put(-15,60){\rotatebox[origin=c]{90}{Bayesian GLM}}
  \put(0,60){\rotatebox[origin=c]{90}{(joint model)}}
\end{picture} &
\includegraphics[height=1.8in]{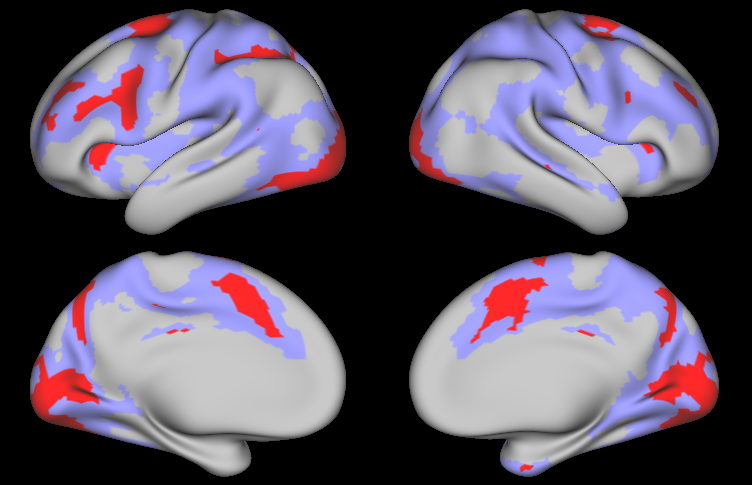}&
\includegraphics[height=1.8in]{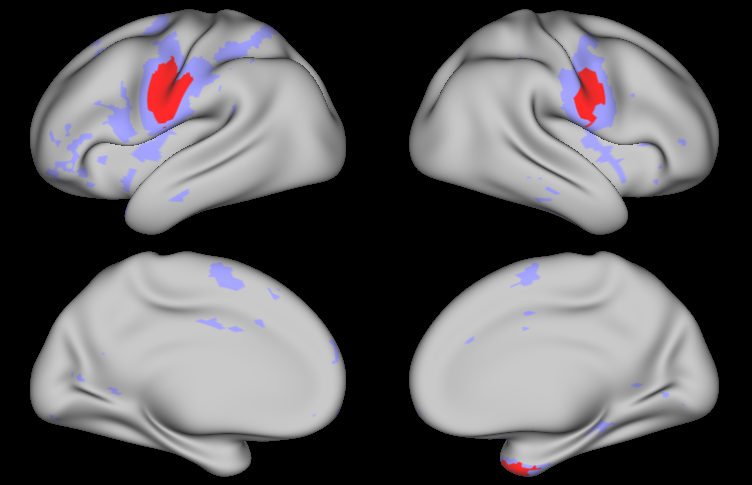}\\[4pt]
\begin{picture}(5,120)
  \put(-15,60){\rotatebox[origin=c]{90}{Bayesian GLM}}
  \put(0,60){\rotatebox[origin=c]{90}{(two-level model)}}
\end{picture} &
\includegraphics[height=1.8in]{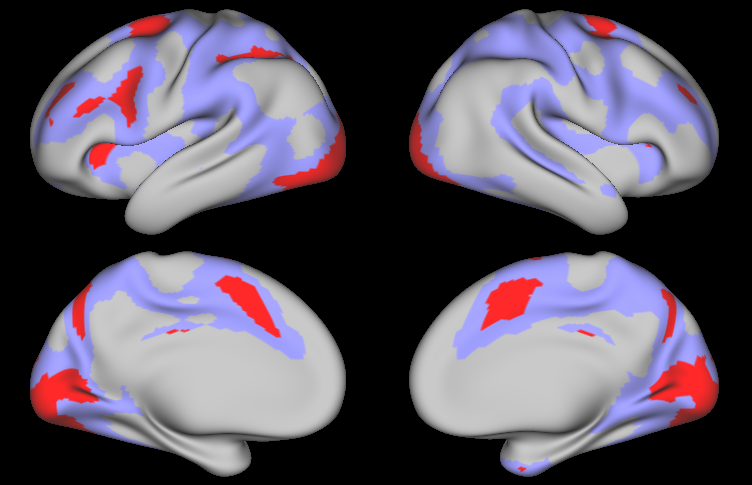}&
\includegraphics[height=1.8in]{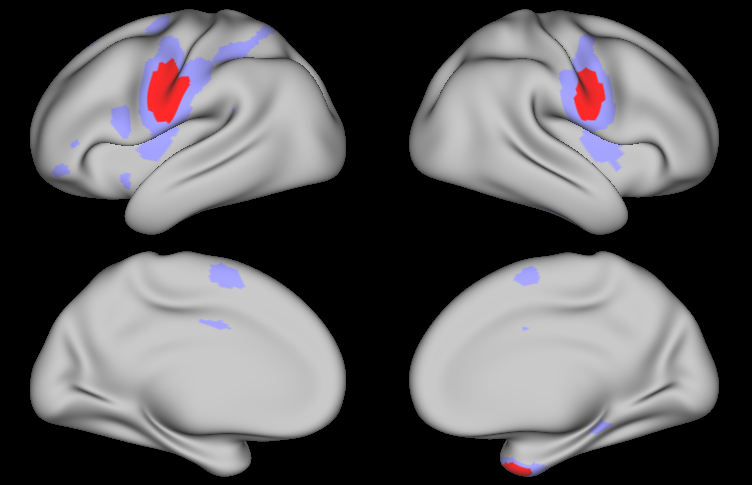}\\[-14pt]
\end{tabular}
\end{center}
\singlespace\caption{\small Group-level regions of activation at significance level $0.01$, as estimated by FWER and FDR control for the classical GLM and the joint PPM excursions set method for the Bayesian GLM.  For the Bayesian GLM, regions of activation at activation threshold $\gamma=0$ are shown in purple, while those at $\gamma=0.027$, corresponding to $1\%$ of the baseline signal, are shown in red.}
\label{fig:app99}
\end{figure}

Figure \ref{fig:app99} shows the regions of activation identified by thresholding the joint PPM excursion function for each task, along with those identified through the classical GLM with FDR and FWER correction, using significance level $0.01$ in each method.  For the Bayesian GLM, regions of activation at activation threshold $\gamma=0$ are shown in purple, while those at $\gamma=0.027$, corresponding to $1\%$ of the global baseline signal, are shown in red.  Several effects are clearly notable.  First, the areas of activation based on controlling the FWER in the classical GLM are very conservative.  This is a well-known issue with this approach that worsens as the number of locations increases, which is likely why FWER correction produces reasonable results in the simulation study but very conservative results on real fMRI data.  FDR control results in significantly larger areas of activation, reflecting an increase in power while maintaining a known degree of false positive control.  Second, the areas of activation based on $\gamma=0$ in the Bayesian GLM are similar to those based on FDR control in the classical GLM, but are smoother and somewhat larger.  This likely reflects an increase in power resulting from incorporating spatial dependencies in the Bayesian GLM, as observed in our simulation study. Third, compared with the joint model, the areas of activation based on the Bayesian two-level model are smoother but similar in size.  This illustrates that while the two-level approach tends to result in somewhat oversmoothed estimates, the sampling method described in Section \ref{sec:2level} is able to appropriately account for uncertainty in the subject-level estimates when fitting the group-level model.  Finally, the increased power in the joint PPM results in widespread areas of activation based on $\gamma=0$, while the more biologically meaningful threshold of $\gamma=0.027$ ($1\%$ of the baseline signal) results in more conservative areas of activation that correspond well to the highly activated areas shown in yellow in Figure \ref{fig:app:beta}.

These results illustrate the benefits of using a Bayesian framework to account for spatial dependencies in fMRI task activation studies.  Compared with the traditional analysis techniques, the proposed Bayesian GLM approach results in smoother estimates of activation for individual subjects and groups of subjects, smoother active regions, and greater power to detect areas of activation, particularly more subtle activations.

\section{Discussion}\label{sec:dis}

In this article, we have proposed a novel Bayesian GLM approach for analysis of cortical surface fMRI (cs-fMRI) data, which has recently experienced a rise in popularity compared to traditional, volumetric fMRI. While cs-fMRI offers several advantages over volumetric fMRI, most analyses of cs-fMRI still utilize the classical ``massive univariate'' GLM approach, which suffers from several pitfalls including dependence on naive smoothing methods and multiple comparisons correction techniques that may not provide accurate control over false positive rates.  By contrast, the Bayesian GLM approach that we propose achieves appropriate smoothing and control of false positives in a model-based fashion and is, to our knowledge, the first spatial Bayesian method proposed for this type of data. The advantages of the proposed Bayesian approach in comparison with the classical GLM have been demonstrated through a simulation study and a motor task fMRI study from the Human Connectome Project.

In comparison with previously proposed Bayesian methods for volumetric fMRI data, the proposed Bayesian approach for cs-fMRI data offers several advantages.  First, the assumptions of isotropy and stationarity are much more reasonable for cs-fMRI data, since geodesic distances along the cortex are more biologically meaningful than Euclidean distances within the volume. Second, a common computational strategy in Bayesian models for volumetric fMRI is to fit a separate model within each slice, an approach that has been shown to result in discontinuities in estimated amplitude maps and regions of activation \citep{siden2017fast}.  Since cs-fMRI data is smaller than volumetric fMRI data and can be resampled in a principled way to further reduce dimensionality with minimal loss of information, it is feasible to fit a single model to all the locations within each hemisphere.  Third, while most Bayesian methods for volumetric fMRI reduce computational burden by using VB, which is known to underestimate posterior variance, we employ INLA, a computationally efficient but highly accurate approximate Bayesian inference tool. Because INLA is less computationally demanding than MCMC, we are able to fit a complex model based on flexible SPDE spatial processes in order to more accurately capture the unique spatial dependency patterns of each latent field.  Fourth, while previously proposed Bayesian methods use the marginal posterior distribution to identify regions of activation, we adopt a computationally efficient excursions set method that uses the joint posterior distribution to account for spatial dependencies in the posterior probabilities of activation.  Finally, we introduce a novel approach for multi-subject analysis, which avoids the computational burden of fitting multiple subjects simultaneously, while providing proper group-level spatial Bayesian inference, a gap in the existing literature \citep{siden2017fast}. None of these techniques have, to the best of our knowledge, ever been proposed for analysis of cortical surface fMRI data.

Several limitations of the proposed methods should be noted.  First,  the SPDE priors we propose in this paper are suitable for modeling stationary and isotropic spatial processes.  In volumetric fMRI data, it has been well-established that amplitude fields tend to present some non-stationary features such as varying degrees of activation \citep[e.g.,][]{harrison08,yue:loh:lind:fmri:10}.  While we largely avoid these issues through the use of cortical surface fMRI data, non-stationary features may exist due to the presence of distinct functional cortical areas and varying degress of smoothness within activated and non-activated areas.  
 \cite{lind:rue:11} introduced non-stationarity to the SPDE models by allowing $\kappa$ and $\tau$ in (\ref{spde:exact}) to depend on location, and \cite{bolin:aoas11} and \cite{fuglstad2013exploring} generalized further using nested directional operators and non-isotropic Laplacians, respectively.  Future work will investigate the feasibility and performance of non-stationary SPDE models in the Bayesian GLM framework.
 
Second, while the joint PPM approach described in Section \ref{sec:ppm} allows us to identify a subset of jointly activated locations, this subset may include isolated regions containing only a single voxel or vertex, which are unlikely to be truly activated in isolation. These could be excluded from the excursion set, thereby guaranteeing that the remaining regions have at least the target activation probability.  However, this would result in overly conservative regions of activation.  Future work should aim to develop methods to identify spatially connected regions of activation based on the joint PPM.  

Finally, while the proposed methods assume that spatial dependence is inversely related to distance, a more complex network structure involving long-range connections may exist in the data.  Future work will incorporate such long-range dependencies between different brain regions during task activity. We also plan to investigate the potential of an empirical Bayesian framework employing estimates of long-range and short-range dependence to reduce the computational burden compared with a fully Bayesian approach such as the one proposed in this paper.  Additionally, the distances utilized in this paper are radial distances along the spherical surface. While the spherical representation of the cortical surface of each hemisphere provides a simple geometric setting while minimizing distance distortions, some level of distortion is inevitable. Future work will focus on utilizing geodesic distances along the true cortical surface of each subject to more accurately capture distance-related spatial dependencies.

\bigskip
\renewcommand{\baselinestretch}{1.0}
\tiny\normalsize
\bibliography{yuyuestats}
\bibliographystyle{sinica}

\newpage

\end{document}